\documentstyle[aps,multicol,epsfig,prl]{revtex}

\begin{document}
 \input epsf
\draft \preprint{HEP/123-qed}

\title{Investigation of the incremental response of soils using
       a discrete element model}
\author{F. Alonso-Marroqu\'{\i}n and H. J. Herrmann}
\address{ ICA1, University of Stuttgart, 
Pfaffenwaldring 27, 70569 Stuttgart, Germany}
\date{\today}
\maketitle


\begin{abstract}

The incremental stress-strain relation of dense packings of
polygons is investigated here by using molecular dynamics
simulations. The comparison of the simulation results to
the continuous theories is performed using explicit expressions 
for the averaged stress and strain over a representative
volume element. The discussion of the incremental response
raises two  important questions of soil deformation:  
Is the incrementally non-linear  theory  appropriate to describe 
the soil mechanical response? 
Does a purely elastic regime exists in the deformation of 
granular materials?. In both cases  our answer will be "no".
The question of stability is also discussed in terms of the
Hill condition of stability for non-associated materials.

\end{abstract}

\begin{multicols}{2}
\section{Introduction}
\label{Intro}

For many years the study of the mechanical behavior of soils was developed in 
the framework of linear elasticity \cite{landau86} and the Mohr-Coulomb 
failure criterion \cite{vermeer98} However, since the start of the development 
of the non-linear constitutive relations in 1968 \cite{roscoe68}, a great 
variety of constitutive models describing different aspects of soils have
been proposed \cite{gudehus84}. A crucial question has been brought forward: 
What it the most appropriate constitutive model to interpret the experimental 
result, or to implement a finite element code? Or more precisely, why is the 
constitutive relation I am using better than that one of the fellow next lab?

In the last years, the discrete element approach has been used as a tool to 
investigate  the mechanical response of soils at the grain level 
\cite{cundall79}. Several average procedures have been proposed to 
define the stress  \cite{bagi96,cundall82,goldenberg02} 
and the strain tensor \cite{bagi99,laetzel02} in terms of the contact forces 
and displacements at the individual grains. These methods have been used to 
perform a direct calculation of the incremental stress-strain relation of 
assemblies of disks \cite{bardet94b} and spheres \cite{kishino03}, without 
any a-priori hypothesis about the constitutive relation. 
Some of the results lead to the conclusion that  the non-associated 
theory of elasto-plasticity is sufficient to describe the observed 
incremental behavior \cite{bardet94b}. 
However, some recent investigations using three-dimensional 
loading paths of complex loading histories seem to contradict 
these results \cite{calvetti03,kishino03}. 
Since the simple spherical geometries of the grains overestimate the role 
of rotations in realistic soils \cite{calvetti03}, it is interesting 
to evaluate the incremental response using arbitrarily shaped particles.

In this paper we investigate the incremental response in the 
quasi-static deformation of dense assemblies of polygonal particles.
The comparison of the numerical simulations with the
constitutive theories is performed by introducing the concept of
{\it Representative Volume Element } (RVE). This volume is chosen the smear 
out the strong fluctuations of the stress and the deformation in the 
granular assembly. In the averaging, each grain is regarded as a piece 
of continuum. By supposing that the stress and the strain of the grain
are concentrated at the small regions of the contacts,  we obtain 
expressions for the averaged stress and strain over the RVE, in terms 
of the contact forces, and the  individual displacements and rotations 
of the grains. The details of this homogenization method are 
presented in Sec. \ref{homogenization}. A short review of the incremental, 
rate-independent stress-strain models is presented in Sec. 
\ref{incremental}. We make special emphasis in the classical 
Drucker-Prager elasto-plastic models and 
the recently elaborated theory of hypoplasticity.
The details of the particle model are presented  in 
Sec. \ref{model}.  The interparticle forces include elasticity, 
viscous damping and   friction with the possibility of slip.
The system is driven by applying stress controlled tests
on a rectangular framework consisting of four walls.
Some loading programs were implemented in Sec. \ref{simulation},
in order to lead to four basic question on the incremental
response of soils: 1) The existence of tensorial zones in the
incremental response, 2) the validity of the superposition
principle and 3) the existence of a finite elastic regime and
4) the question of stability according to the Hill condition.

\section{Homogenization}
\label{homogenization}

The aim of this section is to calculate the macro-mechanical quantities, 
the stress and strain tensors, from micro-mechanical variables of the
granular assembly such as  contact forces, rotations and displacements of 
individual grains.

A particular feature of granular materials is that both the stress and 
the deformation gradient are very concentrated in small regions around 
the contacts between the grains, so that they vary strongly on short 
distances.  The standard homogenization procedure smears out these 
fluctuations by averaging these quantities over a RVE. 
The diameter $d$ of the RVE must be such that $\delta \ll d \ll D$, 
where $\delta$ is the characteristic diameter of the particles and 
$D$ is the characteristic length of the continuous variables.  

\begin{figure}[t]
  \begin{center}
    \epsfig{file=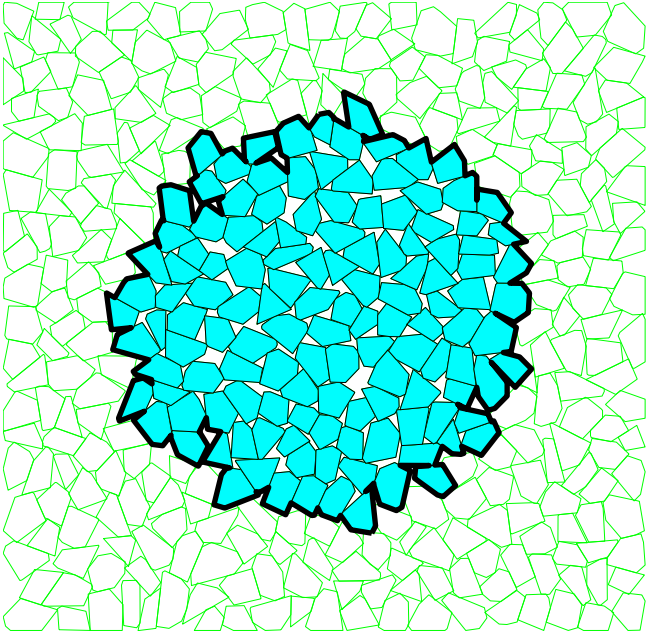,width=\linewidth,clip=1}
    \caption{Representative volume element (RVE).}
    \label{fig:rve}
  \end{center}
\end{figure}

We use here this procedure to obtain the averages of the stress and 
the strain tensors over a RVE in granular materials, which will allow 
us to compare the molecular dynamics simulations to the constitutive 
theories. We regard stress and strain to be continuously distributed 
through the grains, but concentrated at the contacts.
It is important to comment that this averaging procedure 
would not be appropriate to describe the structure of the chain forces 
or the shear band because typical variations of the stress corresponds 
to few particle diameters. Different averaging procedures using 
coarse-grained functions  \cite{goldenberg02}, or cutting the space in 
slide-shaped areas \cite{oda00,laetzel02}, can deal with the question 
of how one can perform averages, and at the same time maintain these 
features.

We will calculate the averages around a point $\vec{x}_0$ of the granular sample 
taking a RVE calculated as follows: at the initial configuration, we 
select the grains whose center of mass are less than $d$ from $\vec{x}_0$. 
Then the RVE is taken as the volume $V$ enclosed by the initial 
configuration of the grains. See Fig. \ref{fig:rve}. 
The diameter $d$ is taken, so that the averaged quantities are 
not sensible to the increase of the diameter  by one particle diameter. 

\subsection{Micro-mechanical stress}

The Cauchy stress tensor is defined using the force acting on an area 
element situated on or within the grains. Let $\vec{f}$ be the force 
applied on a surface element $a$ whose normal unit vector is $\vec{n}$. 
Then the  stress is defined as the tensor satisfying \cite{landau86}:

\begin{equation}
  \sigma_{kj}n_k = lim_{a\to 0}{f_j/a},
\label{cauchy}
\end{equation}

\noindent
where the Einstein summation convention is used. In absence of body 
forces, the equilibrium equations in every volume element lead to 
\cite{landau86}:

\begin{equation}
\partial\sigma_{ij}/\partial x_i = 0.
\label{cauchy1}
\end{equation}

We are going to calculate the average of the stress tensor  
$\bar\sigma$ over the RVE:

\begin{equation}
\bar\sigma = \frac{1}{V}\int_V{\sigma dV}.
\label{sigma}
\end{equation}

Since there is no stress in the voids of the granular media, the 
averaged stress can be written as the sum of integrals taken over 
the particles

\begin{equation}
\bar\sigma = \frac{1}{V} \sum^N_{\alpha=1} {\int_{V_{\alpha}}{\sigma_{ij} dV}},
\label{cauchy2}
\end{equation}

\noindent
where $V_{\alpha}$ is the volume of the particle $\alpha$ and 
$N$ is the number of particles of the RVE. Using the identity 

\begin{equation}
  \frac{\partial (x_i\sigma_{kj})}{\partial x_k} 
 = x_i \frac{\partial\sigma_{kj}}{\partial x_k} + \sigma_{ij},
 \label{cauchy3}
\end{equation}

\noindent
Eq. (\ref{cauchy1}), and the Gauss theorem, Eq. (\ref{cauchy2}) leads to

\begin{equation}
\bar\sigma_{ij}
= \frac{1}{V} \sum_{\alpha} {\int_{V_{\alpha}}{ \frac {\partial (x_i \sigma_{kj} )}{\partial x_k} dV}} 
= \frac{1}{V} \sum_{\alpha} {\int_{\partial V_{\alpha}}{ x_i \sigma_{kj} n_k da}}.
\label{cauchy4}
\end{equation}

The right hand side is the sum over the surface integrals of each grain. 
$\partial V_{\alpha}$ represents the surface of the grain $\alpha$ and 
$\vec{n}$ is the unit normal vector to the surface element $da$.

An important feature of granular materials is that the stress acting on 
each grain boundary is concentrated in the small regions near to the contact 
points. Then we can use the definition given in Eq. (\ref{cauchy}) to express 
this stress on particle $\alpha$ in terms of the contact forces by introducing Dirac delta 
functions:

\begin{equation}
\sigma_{kj} n_k= \sum^{N_{\alpha}}_{\beta=1} {f^{\alpha\beta}_j \delta(\vec{x}-\vec{x}^{\alpha\beta})},
\label{cauchy5}
\end{equation}

\noindent
where $\vec{x}^{\alpha\beta}$ and $\vec{f}^{\alpha\beta}$ are the position and 
the force at the contact $\beta$, and $N_{\alpha}$ is the number of contacts 
of the particle $\alpha$. Replacing Eq. (\ref{cauchy5}) into 
Eq. (\ref{cauchy4}), we obtain

\begin{equation}
\bar\sigma_{ij}= \frac{1}{V}\sum_{\alpha\beta}{x^{\alpha\beta}_i}f^{\alpha\beta}_j. 
\label{cauchy6}
\end{equation}

Now we decompose $\vec{x}^{\alpha\beta}=\vec{x}^{\alpha}+\vec{\ell}^{\alpha\beta}$ 
where $\vec{x}^{\alpha}$ is the position of the center of mass and 
$\vec{\ell}^{\alpha\beta}$ is the branch vector, connecting the center of mass 
of the particle to the point of application of the contact force. 
Imposing this decomposition in Eq. (\ref{cauchy6}), and using the 
equilibrium equations at each particle 
$\sum_{\beta}{\vec{f}^{\alpha\beta}}=0$ we have

\begin{equation}
\bar\sigma_{ij}= \frac{1}{V}\sum_{\alpha\beta}{\ell^{\alpha\beta}_i}f^{\alpha\beta}_j. 
\label{stress1}
\end{equation}

From the equilibrium equations of the torques $\sum_{\beta}(\ell^{\alpha\beta}_i 
f^{\alpha\beta}_j  - \ell^{\alpha\beta}_j f^{\alpha\beta}_i )=0$ one obtains 
that this tensor is symmetric, i. e., 

\begin{equation}
\bar\sigma_{ij}-\bar\sigma_{ji} = 0.
\label{stress2}
\end{equation}

Then, the eigenvalues of this matrix are always real. 
This property leads to some simplifications, as we will see later.


\subsection{Micro-mechanical strain}

In elasticity theory, the strain tensor is defined as the symmetric part of 
the average of the displacement gradient with respect to the equilibrium 
configuration of the assembly. Using the law of conservation of energy, 
one can define the stress-strain relation in this theory \cite{landau86}. 
In granular materials, it is not possible to define the strain in this sense, 
because any loading involves a certain amount of plastic deformation at the 
contacts, so that it is not possible to define the initial reference state to 
calculate the strain. Nevertheless, one can define a strain 
tensor in the incremental sense. This is defined as the average of the 
displacement tensor taken from the deformation during a certain time
interval. 

At the micro-mechanical level, the deformation of the granular materials is 
given by a displacement field $\vec{u}=\vec{r'}-\vec{r}$ at each point of 
the assembly. Here $\vec{r}$ and $\vec{r'}$ are the positions of a material 
point before and after deformation. The average of the strain and rotational 
tensors are defined as:

\begin{equation}
  \label{eq:strain1}
  \bar\epsilon = \frac{1}{2}(F+F^T),
\end{equation}

\begin{equation}
  \label{eq:strain2}
  \bar\omega  = \frac{1}{2}(F-F^T).
\end{equation}

\noindent
Here $F^T$ is the transpose of the deformation gradient $F$, 
which is defined as

\begin{equation}
  \label{eq:deformation01}
   F_{ij}=\frac{1}{V}\int_V{\frac{\partial u_i}{\partial x_j} dV}.  
\end{equation}
  
Using the Gauss theorem, we transform it into an integral over the surface of the RVE

\begin{equation}
  \label{eq:deformation02}
   F_{ij}=\frac{1}{V}\int_{\partial V}{ u_i n_j da},  
\end{equation}

\noindent
where $\partial V$ is the boundary of the volume element. We express this 
as the sum over the boundary particles of the RVE

\begin{equation}
 \label{eq:deformation03}
  F_{ij}=\frac{1}{V}\sum^{N_b}_{\alpha=1}
         \int_{\partial V_{\alpha}}{ u_i n_j da},  
\end{equation}

\noindent
where $N_b$ is the number of boundary particles.
To go further it is convenient to make some approximations. First, the 
displacements of the grains during deformation can be considered rigid 
except for the small deformations near to the contact that can be neglected.
Then, if the displacements are small in comparison to the size of the 
particles, we can write the displacement of the material points inside
of particle $\alpha$ as:

\begin{equation}
  \label{eq:field}
  u_i(\vec{x}) \approx 
\Delta x^{\alpha}_i+e_{ijk}\Delta \phi^{\alpha}_j (x_k - x^{\alpha}_k),
\end{equation}

\noindent
where $\Delta \vec{x}^{\alpha}$, $\Delta \vec{\phi}^{\alpha}$ and 
$\vec{x^\alpha}$ are displacement, rotation and center of mass 
of the particle $\alpha$ which contains the material point $\vec{x}$, 
and $e_{ijk}$ is the anti-symmetric unit tensor.  Setting a parameterization for each 
surface of the boundary grains over the RVE, the deformation gradient 
can be explicitly calculated in terms of grain rotations and 
displacements by replacing Eq. (\ref{eq:field}) in 
Eq. (\ref{eq:deformation03}). 

In the particular case of a bidimensional assembly of 
polygons, the boundary of the RVE is given by a graph   
$\{\vec{x_1}..\vec{x_2},...,\vec{x}_{N_b+1}=\vec{x_1}\}$
consisting of all the edges belonging to the external contour
of the RVE, as shown in Fig. \ref{fig:rve}. 
In this case, Eq. (\ref{eq:deformation03}) can be transformed as 
a sum of integrals over the segments of this contour.

\begin{equation}
  \label{eq:deformation1}
   F_{ij}=\frac{1}{V}\sum^{N_b}_{\beta=1}
{\int^{x_{\beta+1}}_{x_\beta} 
{[\Delta x^\beta_i+e_{ik}\Delta \phi^\beta(x_k- x^\beta_k)] n^{\beta}_j ds}},  
\end{equation}

\noindent
where $e_{ik} \equiv e_{i3k}$ and the unit vector $\vec{n}^{\beta}$ is 
perpendicular to the segment $\overrightarrow{x^{\beta}x^{\beta+1}}$.  
Here $\beta$ corresponds to the index of the boundary segment.
$\Delta \vec{x}^{\beta}$, $\Delta \vec{\phi}^{\beta}$ and 
$\vec{x^\beta}$ are displacement, rotation and center of mass 
of the particle which contains this segment.
Finally, by using the parameterization  
$\vec{x} =\vec{x}^{\beta}+s(\vec{x}^{\beta+1}-\vec{x}^{\beta})$,  
where $(0<s<1)$, we can integrate Eq. (\ref{eq:deformation1}) to obtain

\begin{equation}
F_{ij} = \frac{1}{V}\sum_{\beta}{(\Delta x^\beta_i 
           + e_{ik}\Delta \phi^\beta \ell^{\beta}_k) N^{\beta}_{j}},
\end{equation}

\noindent
where $N^{\beta}_j = e_{j,k}(x^{\beta+1}_k -x^{\beta}_k)$ and 
$\vec{\ell}=(\vec x^{\beta+1} -\vec x^{\beta})/2 - \vec x^\alpha$. The 
stress tensor can be calculated taking the symmetric part of this tensor
using Eq. (\ref{eq:strain1}). 
Contrary to the strain tensor calculated for spherical particles 
\cite{rothenburg88}, the individual rotation of the particles appears 
in the calculation of this tensor. This is given by the fact that for 
non-spherical particles the branch vector is not perpendicular to the 
contact normal vector, so that $e_{ik} \ell^{\beta}_k N^{\beta}_j\ne 0$.

\section{Incremental theory}
\label{incremental}

Since the stress and the strain tensor are symmetric, it is advantageous 
to simplify the notation by defining these quantities as six-dimensional 
vectors:

\begin{equation}
\label{ssv}
\tilde\sigma=\left[ \begin{array}{c}  
\bar\sigma_{11}\\  \bar\sigma_{22}\\ \bar\sigma_{33}\\ 
\sqrt{2} \bar\sigma_{23}\\ \sqrt{2}\bar\sigma_{31}\\ \sqrt{2}\bar\sigma_{13}\\
\end{array} \right]
,~~~ and ~~~~
\tilde\epsilon=\left[ \begin{array}{c}  
\bar\epsilon_{11}\\  \bar\epsilon_{22}\\ \bar\epsilon_{33}\\  
\sqrt{2}\bar\epsilon_{23}\\ \sqrt{2}\bar\epsilon_{31}\\ 
\sqrt{2}\bar\epsilon_{13}\\
\end{array} \right]
\end{equation}

The  coefficient$\sqrt{2}$  allows us to preserve the metric in this
transformation: 
$\tilde\sigma_{k}\tilde\sigma_{k}=\bar\sigma_{ij}\bar\sigma_{ij} $ .
The relation between these two vectors will be established in the 
general context  of the rate-independent incremental constitutive 
relations. We will focus on two particular theoretical developments: 
the theory of hypoplasticity and the elasto-plastic models. The similarities 
and differences of both formulations  are presented in the framework 
of the incremental theory as follows.

\subsection{General framework}
\label{general framework}

In principle, the mechanical response of granular materials can be 
described by a functional dependence of the stress $\tilde\sigma(t)$ 
at time $t$ on  the strain history $\{\tilde\epsilon(t')\}_{0<t'<t}$. 
However, the  mathematical description of this dependence turns out 
to be very  complicated due to the non-linearity and irreversible behavior of 
these materials. An incremental relation, relating the incremental stress 
$d\tilde\sigma(t)=\sigma'(t)dt$ with the incremental strain 
$d\tilde\epsilon(t)=\epsilon'(t)dt$ and some internal variables 
$\chi$ accounting for the deformation history, enables us to avoid 
these mathematical difficulties \cite{darve00}.  This incremental 
scheme is also useful to solve geotechnical problems, since the 
finite element codes require that the constitutive relation be 
expressed incrementally. 

Due to the large number of existing incremental relations, the 
necessity of a unified theoretical framework has been pointed 
out as an essential necessity to classify all the existing models 
\cite{gudehus79} In general, the incremental stress is related to 
the incremental strain by the following function:

\begin{equation}
{\cal F}_\chi (d\tilde\epsilon,d\tilde\sigma,dt).
\label{ce}
\end{equation}

Let's look at the special case where there is no rate dependence 
in the constitutive relation. This means that this relation is not 
influenced by the time required during any loading tests, as corresponds 
to the quasi-static approximation. In this case  ${\cal F}$ is invariant 
with respect to $dt$, and Eq. (\ref{ce}) can be reduced to:

\begin{equation}
d\tilde{\epsilon}={\cal G}_\chi(d\tilde\sigma)
\label{dce}
\end{equation}

In particular, the rate-independent condition implies that multiplying 
the loading time by a scalar $\lambda$ does not affect the incremental 
stress-strain relation:

\begin{equation}
\forall \lambda,~~~{\cal G}_\chi(\lambda d\tilde\sigma) 
=\lambda{\cal G}_\chi(d\tilde\sigma)
\label{rti}
\end{equation}
 
This equation means that ${\cal G}_\chi$ is an homogeneous function of 
degree one. In this case, the application of the Euler identity shows 
that Eq. (\ref{dce}) leads to 

\begin{equation}
d\tilde{\epsilon}=M_\chi(\hat\sigma)d\tilde\sigma,
\label{eq:incremental}
\end{equation}

\noindent
where $M_\chi = \partial{\cal G}_\chi/\partial(d\tilde\sigma)$ and 
$\hat\sigma$ is the unitary vector defining the direction of the 
incremental stress

\begin{equation}
\hat{\sigma}= \frac{d\tilde{\sigma}}{|d\tilde{\sigma}|}.
\label{dir}
\end{equation}

Eq. (\ref{eq:incremental}) represents the general expression for the 
rate-independent constitutive relation. In order to determine the 
dependence of $M$ on $\hat\sigma$, one can either perform 
experiments by taking different loading directions, or postulate 
explicit expressions based on a theoretical framework. The first 
approach will be considered in the next section, and the discussion 
about some existing theoretical models will be presented as follows.

\subsection{Elasto-plastic models}
\label{elasto-plastic}

The classical theory of elasto-plasticity has been established by 
Drucker and Prager in the context if metal plasticity \cite{drucker52}. 
Some extensions have been developed to describe soils, clays, rocks, 
concrete, etc. \cite{vermeer98,nova79}.  Here we  present 
a short review of these developments in soil mechanics. 

\begin{figure}[htbp]
  \begin{center}
    \begin{minipage}{0.5\linewidth}
      \begin{center}{ {\bf (a)} }
        \parbox[l]{\linewidth}{
          \includegraphics[width=\linewidth,clip=1]{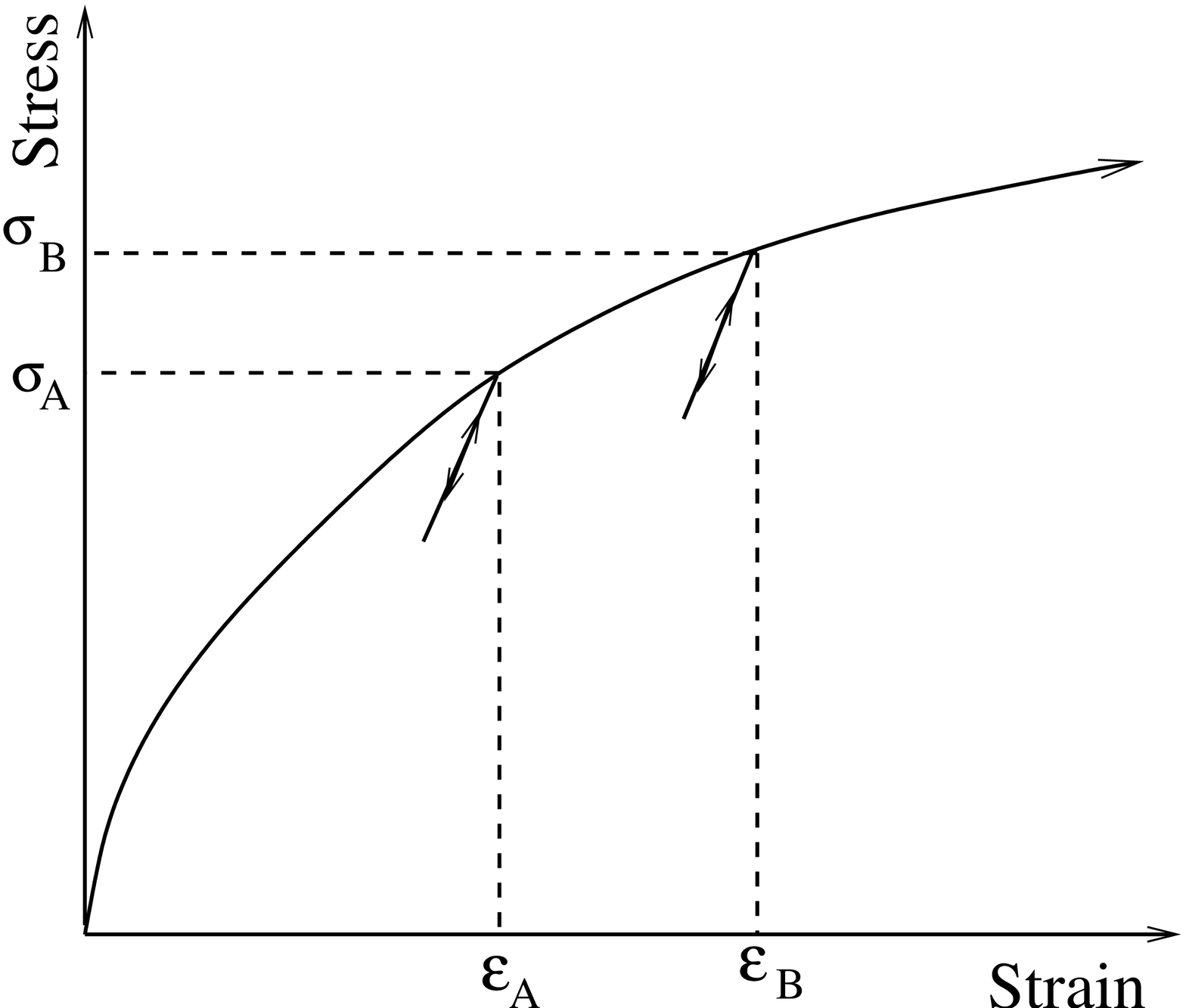}
          }
      \end{center}
    \end{minipage}
    \begin{minipage}{0.48\linewidth}
      \begin{center} {{\bf (b)} }
        \parbox[c]{\linewidth}{
          \includegraphics[width=\linewidth,clip=1]{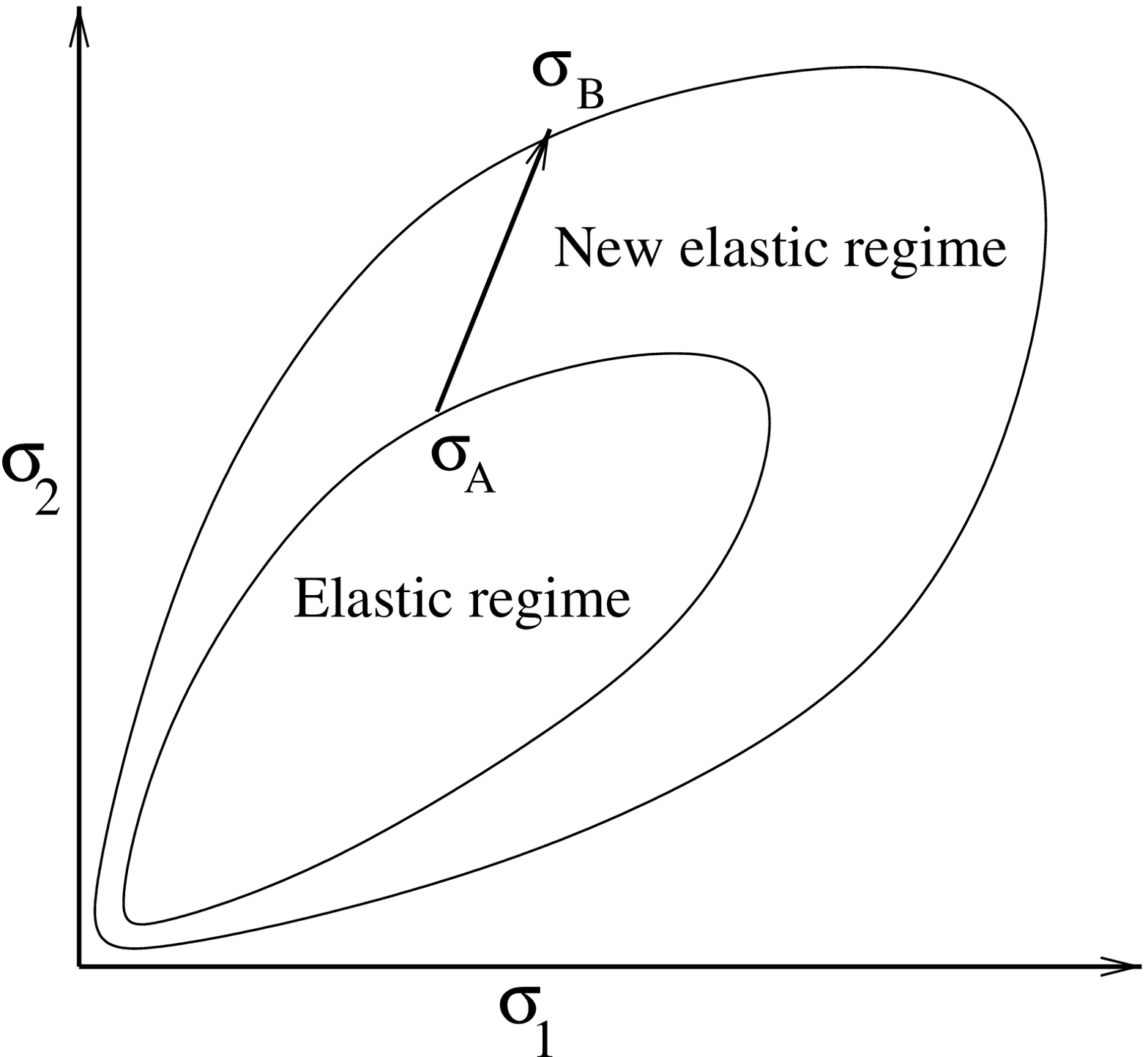}
          }
      \end{center}
    \end{minipage}
    \caption{ Evolution of the elastic regime {\bf a)}
stress-strain relation {\bf b)} elastic regime in the stress space.}
    \label{fig:plastic}
  \end{center}
\end{figure}

When a granular sample subjected to a confining pressure is loaded in the 
axial direction, it displays a typical stress-strain response as shown 
in the left part of Fig. \ref{fig:plastic}. A continuous decrease of the 
stiffness (i.e. the slope of the stress-strain curve) is observed 
during the loading. If the sample is unloaded, an abrupt increase 
in the stiffness is observed, leaving an irreversible 
deformation. One observes that if the stress is changed around 
some region below  $\sigma_A$, which is called the 
{\it yield point}, the deformation is almost linear and 
reversible. The first postulate of the elasto-plastic theory 
establishes a stress region immediately below the yield point 
where only elastic deformations are possible.

{\bf
Postulate 1: For each stage of loading there exists a yield surface, 
which  encloses a finite  region in the stress space where only reversible 
deformations are possible. }

The simple Mohr-Coulomb model assumes that the onset of plastic 
deformation occurs at failure \cite{vermeer98}. However, it has 
been experimentally shown that plastic deformation develops before 
failure \cite{roscoe70}. In order to provide a consistent description 
of these experimental results with the elasto-plastic theory, it is 
necessary to suppose that the yield function changes with the deformation 
process \cite{roscoe70}. 
This condition is schematically shown in Fig. \ref{fig:plastic}.
Let suppose that the sample is loaded until it reaches the stress 
$\sigma_A$  and then it is slightly unloaded. If the sample is reloaded,
it is able to recover the same stress-strain relation of the
monotonic loading once it reaches the yield point $\sigma_A$ again. 
If one increases  the load to the stress  $\sigma_B$, a new elastic 
response can be observed after a loading reversal. In the 
elasto-plasticity context, this result is  interpreted by supposing 
that the elastic regime is expanded to a new  stress region below 
the yield point $\sigma_B$. 

{\bf
Postulate 2: The yield function remains when the deformations take place 
inside of the elastic regime, and it changes as the plastic deformation 
evolves. }

The transition from the elastic to the elasto-plastic response is 
extrapolated for more general deformations. Part (b) of 
Fig. \ref{fig:plastic} shows the evolution of the elastic region 
when the yield point moves in the stress space from $\tilde{\sigma_A}$ 
to $\tilde{\sigma_B}$. The essential goal in the elasto-plastic theory 
is to find the correct description of the evolution of the elastic 
regime with the deformation, which is called the {\it hardening law}.

We will finally introduce the third basic assumption from elasto-plasticity
theory:

{\bf
Postulate 3: The strain can be separated in an elastic (recoverable) and a 
plastic (unrecoverable) component:}

\begin{equation}
d\tilde{\epsilon}  = d\tilde{\epsilon}^e+ d\tilde{\epsilon}^p,
\label{eq:elastic1}
\end{equation}

The incremental elastic strain is linked to the incremental stress by 
introducing an elastic tensor as

\begin{equation}
d\bar{\sigma}= D(\tilde{\sigma})d\tilde{\epsilon}^e.
\label{elastic}
\end{equation}

To calculate the incremental plastic strain, we introduce the  
{\it yield surface} as

\begin{equation}
  f(\sigma,\kappa)=0,
  \label{eq:yield}
\end{equation}

\noindent
where $\kappa$ is introduced as an internal variable, which describes the 
evolution of the elastic regime with the deformation. From experimental 
evidence, it has been shown that this variable can be taken as a function 
of the cumulative plastic strain \cite{vermeer98,nova79} 

\begin{equation}
   \epsilon^p\equiv \int^t_0{\sqrt{d\epsilon_k d\epsilon_k}dt}
\end{equation}

When the stress state reaches the yield surface, the plastic deformation 
evolves. This is assumed to be derived from a scalar function of the stress 
as follows:

\begin{equation}
  \label{eq:potential}
  d\epsilon^p_j=\Lambda \frac{\partial g}{\partial \sigma_j},
\end{equation}

\noindent
where $g$ is the so-called {\it plastic potential} function. 
following the Drucker-Prager postulates it can be shown that $g=f$ 
\cite{drucker52}. However, a considerable amount of experimental
evidence has shown that in soils the plastic deformation is not 
perpendicular to the yield surfaces  \cite{poorooshasb67}. It is 
necessary to introduce this plastic potential to extend the 
Drucker-Prager models to the so-called {\it non-associated} 
models.

The parameter $\lambda$ of Eq. (\ref{eq:potential}) can be obtained from 
the so-called {\it consistence  condition}. This condition comes from 
the second postulate, which establishes that after the movement of the 
stress state from $\tilde{\sigma_A}$ to 
$\tilde{\sigma_B}=\tilde{\sigma_A}+\tilde{d\sigma}$ 
the elastic regime must be expanded so that $df=0$, as shown in Part (b) 
of Fig. \ref{fig:plastic}. Using the chain rule one obtains:

\begin{equation}
  \label{eq:consistente}
  df=\frac{\partial f}{\partial \sigma_i}d\sigma_i
+\frac{\partial f}{\partial\kappa}\frac{\partial\kappa}{\partial\epsilon^p_j} 
     d\epsilon^p_j = 0.
\end{equation}

Replacing Eq. (\ref{eq:potential}) in Eq. (\ref{eq:consistente}), we obtain 
the parameter $\Lambda$

\begin{equation}
  \label{eq:Lambda}
  \Lambda = -
(\frac{\partial f}{\partial \kappa}\frac{\partial \kappa}{\partial\epsilon^p_j}
\frac{\partial g}{\partial \sigma_j})^{-1}
 {\frac{\partial f}{\partial \sigma_i}d\sigma_i}.
\end{equation}

We define the vectors $N^y_i=\partial f/\partial\sigma_i$ and 
$N^f_i=\partial g/\partial\sigma_i$ and the unit vectors  
$\hat\phi=\vec N^y/|\vec{N^y}|$ and $\hat\psi=\vec N^f/|\vec{N^f}|$  as 
the {\it flow direction} and the {\it yield direction}.
In addition, the {\it plastic modulus} is defined as 

\begin{equation}
  \label{eq:hardening1}
  h=-\frac{1}{|\vec{N^y}||\vec{N^f}|} \frac{\partial f}{\partial \kappa}\frac{\partial \kappa}{\partial \epsilon^p_i}
\frac{\partial g}{\partial \sigma_i}.  
\end{equation}

Replacing Eq. (\ref{eq:Lambda}) in Eq. (\ref{eq:potential})
and using Eq. (\ref{eq:hardening1}) we obtain:

\begin{equation}
  \label{eq:plastic-flow1}
  d\tilde\epsilon^p=\frac{1}{h}\hat\phi \cdot d\tilde\sigma ~\hat\psi.
\end{equation}

Note that this equation has been calculated for the case that the stress
increment goes outside of the yield surface. If the stress 
increment takes place inside the yield surface, the second Drucker-Prager 
postulate establishes that $d\tilde\epsilon^p=0$. Thus, the generalization of 
Eq. (\ref{eq:plastic-flow1}) is given by

\begin{equation}
  \label{eq:plastic-flow2}
   d\tilde\epsilon^p=\frac{1}{h}\langle\hat\phi 
   \cdot d\tilde\sigma\rangle ~\hat\psi,
\end{equation}

\noindent
where $\langle x \rangle=x$ when $x>0$ and  $\langle x \rangle =0$ otherwise. 
Finally, the total strain response can be obtained from Eqs. 
(\ref{eq:elastic1}) and (\ref{eq:plastic-flow2}):

\begin{equation}
  \label{eq:elasto-plastic}
  d\epsilon=D^{-1}(\sigma)d\sigma
+\frac{1}{h}\langle\hat\phi \cdot d\tilde\sigma\rangle ~\hat\psi
\end{equation}

From this equation one can distinguish two zones in the incremental stress
space where the incremental relation is linear. They are the so-called 
tensorial zones defined by Darve \cite{darve00}. The existence of two 
tensorial zones, and the continuous transition of the incremental 
response at their boundary, are essential features of the elasto-plastic 
models.

Although the elasto-plastic theory has shown to work well for monotonically 
increasing loading, it has shown some deficiencies in the description 
of complex loading histories \cite{wood82}. 
There is also an extensive body of experimental evidence that shows that 
the elastic regime is extremely small and that the transition from elastic 
to an elasto-plastic response is rather smooth \cite{gudehus84}.  

The concept of {\it bounding surface} has been introduced to generalize 
the classical elasto-plastic concepts \cite{dafalias75}. In this model, 
for any given state within the surface, a proper mapping rule associates 
a corresponding {\it image} stress point on this surface. A measure 
of the distance between the actual and the image stress points is used 
to specify the plastic modulus in terms of a plastic modulus at 
the image stress state. Besides the versatility of this theory to 
describe a wide range of materials, it has the advantage that the elastic 
regime can be considered as vanishingly small, and therefore used to give 
a realistic description of unbound granular soils. 

It is the author's opinion that the most striking aspect of the bounding 
surface  theory with vanishing elastic range is that it establishes a 
convergence point for two different approaches of constitutive modeling: 
the elasto-plastic approaches originated from the Drucker-Prager theory, 
and the more recently developed hypoplastic theories.


\subsection{Hypoplastic models}
\label{hypoplastic}

In recent years, an alternative approach for modeling soil behavior 
has been proposed, which departs from the framework of the elasto-plastic 
theory \cite{kolymbas91,darve95,chambon94}. The distinctive features 
of this approach are:

\begin{itemize}
\item {\bf The absence of the decomposition of strain in plastic and 
elastic components.}

\item{\bf The  statement of a non-linear dependence of the incremental 
response with the strain rate directions.}
\end{itemize}

The most general expression has been provided by the so-called second 
order incremental non-linear models \cite{darve95}. A particular 
class of these models which has received special attention in recent 
times is provided by the theory of hypoplasticity 
\cite{kolymbas91,chambon94}. A general outline of this theory was laid 
down by Kolymbas \cite{kolymbas91}, leading to different formulations, 
such as the K-hypoplasticity developed in Karlsruhe \cite{wu96}, 
and the CLoE-hypoplasticity originated in Grenoble \cite{chambon94}. 
In the hypoplasticity, the most general constitutive equation takes the 
following form:

\begin{equation}
  \label{eq:hypoplastic1}
  d\tilde\sigma = \ L(\tilde\sigma,\nu)d\tilde\epsilon 
                + \tilde N(\tilde\sigma,\nu)|d\tilde\epsilon|,
\end{equation}

\noindent
where $L$ is a second order tensor and $\tilde N$ is a vector, both 
depending on the current state of the material, the stress $\tilde\sigma$ 
and the void ratio $\nu$. Hypoplastic equations provide a simplified 
description of loose and dense unbound granular materials. A reduced 
number of parameters are introduced, which are very easy to calibrate 
\cite{gudehus99}. 

In the theory of hypoplasticity, the stress-strain relation is established  
by means of an incremental non-linear relation without any recourse to yield 
or boundary surfaces. This non-linearity is reflected in the fact
that the relation between the incremental stress and the incremental
strain given in Eq. (\ref{eq:hypoplastic1}) is always non-linear.
The incremental non-linearity of the granular materials is still 
under discussion. Indeed, an important feature of the incremental
non-linear constitutive models is that they break away from the 
superposition principle:

\begin{equation}
  \label{eq:superposition1}
  d\tilde\sigma(d\tilde\epsilon_1+d\tilde\epsilon_2) 
    \ne  d\tilde\sigma(d\tilde\epsilon_1) + d\tilde\sigma(d\tilde\epsilon_2),
\end{equation}

\noindent
which is equivalent to:

\begin{equation}
  \label{eq:superposition2}
  d\tilde\epsilon(d\tilde\sigma_1+d\tilde\sigma_2) 
    \ne  d\tilde\epsilon(d\tilde\sigma_1) + d\tilde\epsilon(d\tilde\sigma_2)
\end{equation}

\begin{figure}[t]
  \begin{center}
    \epsfig{file=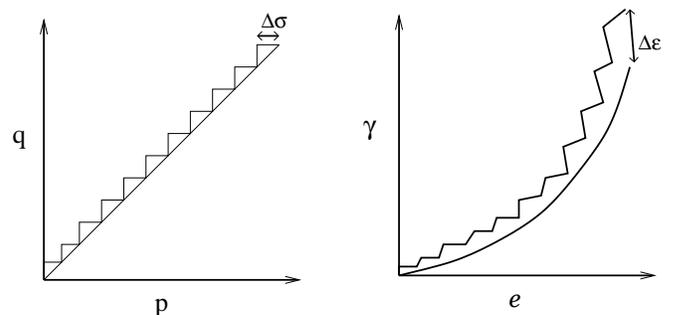,width=\linewidth,clip=1}
    \caption{Smooth and stair-like stress paths and corresponding strain 
responses. $p$ and $q$ represent the pressure and the deviatoric stress.
$e$ and $\gamma$ are the volumetric and deviatoric strain components.}
    \label{fig:superposition}
  \end{center}
\end{figure}

An important consequence of this feature is addressed by Kolymbas 
\cite{kolymbas93} and Darve \cite{darve95}. They consider two stress paths; 
the first one is smooth and the second one results from the superposition of 
small deviations as shown in Fig. \ref{fig:superposition}. The superposition 
principle establishes that the strain response of the stair-like path 
converges to the response of the proportional loading in the limit 
of arbitrarily small deviations. More precisely, the strain deviations 
$\Delta\epsilon$ and the steps of the stress increments $\Delta\sigma$ satisfy 
$\lim_{\Delta\sigma\to 0}{\Delta\epsilon}=0$. For the hypoplastic equations, 
and in general for the incremental non-linear models, this condition is never 
satisfied. For incremental relations with tensorial zones, this principle 
is satisfied whenever the increments take place inside one of these tensorial 
zones. It should be added that there is no experimental evidence disproving 
or confirming this rather questionable superposition principle.

\section{Discrete model}
\label{model}
 
We present here a two-dimensional discrete element model which will be
used to investigate the incremental response of granular materials.
This model consists of randomly generated convex polygons, which 
interact via contact forces.
There are some limitations in the use of such a two-dimensional 
code to model physical phenomena that are three-dimensional in nature. 
These limitations have to be kept in mind in the interpretation of the 
results and its comparison with the experimental data. In order to give 
a three-dimensional picture of this model, one can consider the polygons 
as a collection of prismatic bodies with randomly-shaped polygonal basis. 
Alternatively, one could consider the polygons as three-dimensional grains 
whose centers of mass all move in the same plane.  It is the author's 
opinion that it is more sensible to consider this model as an idealized 
granular material that can be used to check the constitutive models.

The details of the particle generation, the contact forces, the  
boundary conditions and the molecular dynamics simulations are presented 
in this section.

\subsection{Generation of polygons}
\label{voronoi}

\begin{figure}[t]
  \begin{center}
    \epsfig{file=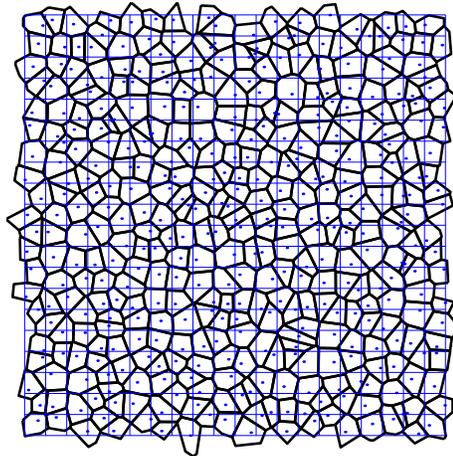,width=0.7\linewidth,clip=1}
    \caption{Voronoi construction used to generate the convex polygons.  
The dots indicate the point used in the tessellation. Periodic boundary 
conditions were used.}
    \label{fig:voronoi}
  \end{center}
\end{figure}

The polygons representing  the particles in this model are generated
by using the method of Voronoi tessellation \cite{kun99}.
This methods is schematically shown in Fig. \ref{fig:voronoi}:
First, a regular square lattice of side $\ell$ is created. Then, we
choose a random point in each  cell of the rectangular grid. Then, each 
polygon is constructed assigning  to  each point that part of the plane 
that is nearer to it than  to any other point. The details of the 
construction of the Voronoi cells can be found in the literature 
\cite{moukarzel92,okabe92}.

Using the Euler theorem, It has been shown analytically that the mean
number of edges of this Voronoi construction must be $6$ \cite{okabe92}. 
The number of edges  of the polygons is distributed between  $4$ and  
$8$ for $98.7\%$ of the polygons. 
It is also found that the orientational  distribution of  edges is 
isotropic, and the distribution of areas of polygons is  symmetric 
around its mean value $\ell^2$.
The probabilistic distribution  of areas  follows approximately a  
Gaussian  distribution with variance  of $0.36\ell^2$.

\subsection{Contact forces}
\label{contact_forces}

In order to calculate the forces, we assume that all the polygons 
have the same thickness $L$. The force between two polygons is 
written as  $\vec{F}=\vec{f} L$ and the mass of the polygons is 
$M=m L$.  In reality, when two elastic bodies come into contact, 
they have a slight deformation in the contact region. In the 
calculation of the contact force we suppose that the polygons 
are rigid, but we allow them to overlap. Then, we calculate 
the force from this virtual overlap.

\begin{figure}
  \begin{center}
     \epsfig{file=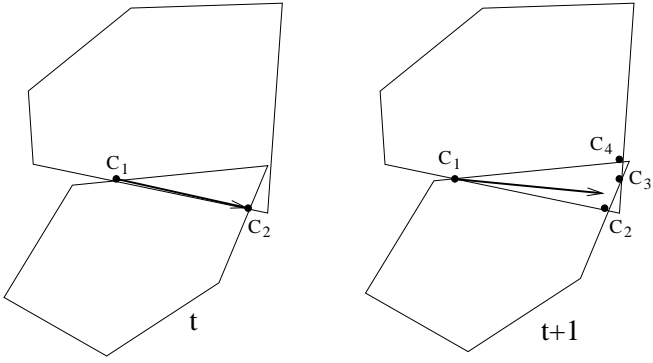,width =\linewidth}
   \end{center}
   \caption{ Contact points $C_i$ before (left) and after the 
formation of a pathological contact (right). The vector denotes 
the contact line. $t$ represents the time step.}
    \label{fig:overlap}
\end{figure}

The first step for the calculation of the contact force is the 
definition of the line representing the flattened contact surface 
between the two bodies in contact.  This is defined from the 
contact points resulting from the intersection of the edges of 
the overlapping polygons. In most  cases, we have two contact 
points, as shown in the left of Fig. \ref{fig:overlap}. 
In such a case, 
the contact line is defined by the vector $\vec{C}=\overrightarrow{C_1 C_2}$ 
connecting these two intersection points. In some pathological
cases, the intersection of the polygons leads to four or six contact 
points. In these cases, we define the contact line by the vector 
$\vec{C}=\overrightarrow{C_1 C_2}+\overrightarrow{C_3 C_4}$ or 
$\vec{C}=\overrightarrow{C_1 C_2}+\overrightarrow{C_3 C_4}+
\overrightarrow{C_5 C_6}$, respectively. 
This choice guarantees a continuous change of the contact line, 
and therefore of the contact forces, during the evolution of the 
contact.

The contact force is separated as $\vec{f}^c=\vec{f}^e+\vec{f}^v$, 
where $\vec{f}^e$  and  $\vec{f}^v$ are the elastic and viscous 
contribution.  The elastic part of the contact force is decomposed 
as $\vec{f^e}= f^e_n \hat{n}^c + f^e_t \hat{t}^c$. The calculation 
of these components is explained below. The unit tangential vector 
is defined as  $\hat{t}^c=\vec{C}/|\vec{C}|$, 
and the normal unit vector $\hat{n}^c$ is taken perpendicular 
to $\vec{C}$. The point of application of the contact force is 
taken as the center of mass of the overlapping polygons.

As opposed to the Hertz theory for round contacts, there 
is no exact way to calculate the normal force between interacting 
polygons. An alternative method has been proposed in order to model 
this force\cite{tillemans95}. In this method, the normal elastic force 
is calculated as $f^e_n= -k_n A/L_c$ where $k_n$ is the normal stiffness,  
$A$ is the overlapping area and $L_c$ is a characteristic length 
of the polygon pair.  
Our choice of $L_c$ is $1/2(1/R_i+1/R_j)$ where 
$R_i$ and $R_j$ are the radii of the circles of the same area as 
the polygons. This normalization is necessary to be consistent 
in the units of force \cite{kun99}. 

In order to model the quasi-static friction force, we calculate 
the elastic tangential force using an extension of the method 
proposed by Cundall-Strack \cite{cundall79}. An elastic force   
$f^e_t= -k_t \Delta x_t $ proportional to the elastic displacement 
is included at each contact. $k_t$ is the tangential stiffness. 
The elastic displacement $\Delta x_t $ is calculated as the time 
integral of the tangential velocity of the contact during the 
time where the elastic condition  $|f^e_t|<\mu f^e_n$ is satisfied. 
The sliding condition is imposed, keeping this force constant when 
$|f^e_t|=\mu f^e_n$. The straightforward calculation of this elastic 
displacement is given by the time integral starting at the beginning 
of the contact:

\begin{equation}
\Delta x^e_t=\int_{0}^{t}v^c_t(t')\Theta(\mu f^e_n-|f^e_t|)dt',
\label{friction} 
\end{equation}

\noindent
where $\Theta$ is the Heaviside step function and $\vec{v}^c_t$ 
denotes the tangential component of the relative velocity $\vec{v}^c$ 
at the contact:

\begin{equation}
\vec{v}^c=\vec{v}_{i}-\vec{v}_{j}-\vec{\omega}_{i}\times\vec{\ell}_{i}
+\vec{\omega}_{j}\times\vec{\ell}_{j}.
\end{equation}

\noindent
Here $\vec{v}_i$ is the linear velocity and $\vec{\omega}_i$ is the 
angular velocity  of the particles in contact. The branch vector 
$\vec{\ell}_i$ connects the center of mass of particle $i$ with 
the point of application of the contact force.

Damping forces are included in order to allow for rapid relaxation 
during the preparation of the sample, and to reduce the acoustic 
waves produced during the loading. These forces are calculated as 

\begin{equation}
\vec{f}^c_v  = -m(\gamma_n v^c_n \hat{n}^c + \gamma_t v^c_t \hat{t}^c),
\label{dm0}
\end{equation}

\noindent
being $m=(1/m_i+1/m_j)^{-1}$  the effective mass of the polygons 
in contact. $\hat{n}^c$ and $\hat{t}^c$ are the normal and tangential 
unit vectors defined before, and $\gamma_n$ and $\gamma_t$ are the 
coefficients of viscosity.  These forces introduce time dependent effects
during the cyclic loading. However, we will show that these effects can 
be arbitrarily reduced by increasing the loading time, as 
corresponds to the quasi-static approximation.

\subsection{Molecular dynamics simulation}
\label{MD}

The evolution of the position $\vec{x}_i$ and the orientation 
$\varphi_i$ of the $i_{th}$ polygon is governed 
by the equations of motion:

\begin{eqnarray}
 m_i\ddot{\vec{x}}_i &=&\sum_{c}\vec{f^c_i}  
+\sum_{b}\vec{f}^b_i, \nonumber\\
I_i\ddot{\varphi}_{i} &=&\sum_{c}\vec{\ell}^c_i\times\vec{f^c_i}
+\sum_{b}\vec{\ell}^b_i\times\vec{f}^b_i. 
\label{dm}
\end{eqnarray}

\noindent
Here $m_i$ and $I_i$ are the mass and moment of inertia of the polygon $i$. 
The first summation goes over all particles in contact with this polygon.
According to the previous section, these  forces  $\vec{f^c}$ are given by

\begin{eqnarray}
\vec{f^c}&=&-(k_n A/L_c + \gamma_n m v^c_n)\vec{n}^c 
-(\Delta x^c_t + \gamma_t m v^c_t)\vec{t}^c, \nonumber \\
\label{dm2} 
\end{eqnarray}
 
\noindent


The second summation on the left hand of Eq. \ref{dm2} goes over all the 
vertices of the polygons in contact with the walls.
This interaction  is modeled by using a simple visco-elastic force. 
First, we allow the polygons to penetrate the walls. Then, for each 
vertex of the polygon $\alpha$ inside of the walls we include a force

\begin{equation}
  \label{box}
  \vec{f}^b= -k_n\delta\vec{n}-\gamma_b m_\alpha \vec{v}^b,
\end{equation}

\noindent
where $\delta$ is the penetration length of the vertex, $\vec{n}$ is 
the unit normal vector to the wall, and $\vec{v}^b$ is the relative 
velocity of the vertex with respect to the wall.
 
We use a fifth-order Gear predictor-corrector method for solving 
the equation of motion \cite{allen87}. This algorithm consists of 
three steps. The first step predicts position and velocity of 
the particles by means of a Taylor expansion. The second step 
calculates the forces as a function of the predicted positions 
and velocities. The third step corrects the positions and
 velocities in order to optimize the stability of the algorithm. 
This method is much more efficient than the simple Euler approach 
or the Runge-Kutta method, especially for problems where very high 
accuracy is a requirement.
 
The parameters of the molecular dynamics simulations were adjusted 
according to the following criteria: 
1) guarantee the stability of the numerical solution,
2) optimize the time of the calculation, and
3) provide a reasonable agreement with the experimental data.
 
There are many parameters in the molecular dynamics algorithm. 
Before choosing them, it is convenient to make a dimensional 
analysis. In this way, we can keep the scale invariance of the model 
and reduce the parameters to a minimum of dimensionless constants.
First, we   introduce the following characteristic times of the 
simulations:  the loading time $t_0$, the relaxation times  
$t_n=1/\gamma_n$, $t_t=1/\gamma_t$, $t_b=1/\gamma_b$ and the 
characteristic period of oscillation  $t_s=\sqrt{k_n/\rho\ell^2}$ 
of the normal contact.

Using the Buckingham Pi theorem \cite{buckingham14}, one can show that the 
strain response, or any other dimensionless variable measuring the response 
of the assembly during loading, depends only on the following dimensionless 
parameters:  $\alpha_1 = t_n/t_s$, $\alpha_2 = t_t/t_s$,  
$\alpha_3 = t_b/t_s$, $\alpha_4= t_0/t_s$, the ratio $\zeta=k_t/k_n$ 
between the stiffnesses, the friction  coefficient $\mu$ and the ratio 
$\sigma_i/k_n$ between the stresses applied on the walls and the normal 
stiffness.  

The variables $\alpha_i$ will be called {\it control parameters}. 
They are chosen in order to satisfy the quasi-static approximation, 
i.e. the response of the system does not depend on the loading time, 
but a change of these parameters within this limit does not affect the 
strain response. $\alpha_2 = 0.1$ and $\alpha_2=0.5$ were taken large 
enough to have a high dissipation, but not too large to keep
the numerical stability of the method.  $\alpha_3 = 0.001$ is chosen 
by optimizing the time of consolidation of the sample in the Subsec.
\ref{loose}. The ratio $\alpha_4=t_0/t_s=10000 $ was chosen large enough 
in order to avoid rate-dependence in the strain 
response, corresponding to the quasi-static approximation.  Technically, 
this is performed by looking for the value of  $\alpha_4$ such that a 
reduction of it by half makes a  change of the stress-strain relation 
less than $5\%$.

The parameters $\zeta$ and $\mu$ can be considered as 
{\it material parameters}.  They determine the constitutive response 
of the system, so they must be adjusted to the experimental data. 
In this study, we have adjusted them by comparing the 
simulation of biaxial tests of dense polygonal packings 
with the corresponding  tests with dense Hostun sand \cite{marcher01}. 
First, the initial Young modulus  of the material is linearly related 
to the value of normal stiffness of the contact. 
Thus, $k_n=160MPa$ is chosen by fitting the initial slope of the stress-strain 
relation in the biaxial test. Then, the Poisson ratio depends on the ratio 
$\zeta=k_t/k_n$. Our choice $k_t=52.8MPa$ gives an initial Poisson ratio 
of $0.07$.  This is obtained from the initial slope of the curve of 
volumetric strains versus  axial strain. The angles of friction and the 
dilatancy are increasing functions  of the friction coefficient $\mu$. 
Taking $\mu=0.25$ yields angles of friction  of $30-40$ degrees and 
dilatancy angles of $10-20$ degrees, which are similar to the 
experimental data of river sand \cite{desrues84}.

\subsection{Sample preparation}
\label{loose}

The Voronoi construction presented above leads to samples 
with no porosity.  This ideal case contrasts with realistic soils, 
where only  porosities up to a certain value can be achieved. In 
this section,  we present a method to create stable, irregular
packings of polygons with a given porosity. 

The porosity can be defined using the concept of void ratio. 
This is defined as the ratio of the volume of the void space 
to the volume of the solid material. It can be calculated as:

\begin{equation}
  \label{eq:void ratio}
  \nu = \frac{V_t}{V_f-V_0}-1. 
\end{equation}

\noindent
This is given in terms of the area enclosed by the floppy 
boundary $V_t$, the sum of the areas of the polygons $V_f$ 
and the sum of the overlapping areas between them $V_0$.

Of course, there is a maximal void ratio that can be 
achieved, because it is impossible to pack particles with an 
arbitrarily high porosity. The maximal void ratio $\nu_m$ 
can be detected by moving the walls until a certain void 
ratio is reached. We find a critical value, above which the 
particles can be arranged without touching. 
Since there is no contacts, the assembly cannot support
a load, and even applying gravity will cause it to compactify.
For a void  ratio below this critical value, there will be particle 
overlap, and the assembly is able to sustain a certain load. 
This maximal value is around $0.28$.

\begin{figure}[b]
  \begin{center}
    \epsfig{file=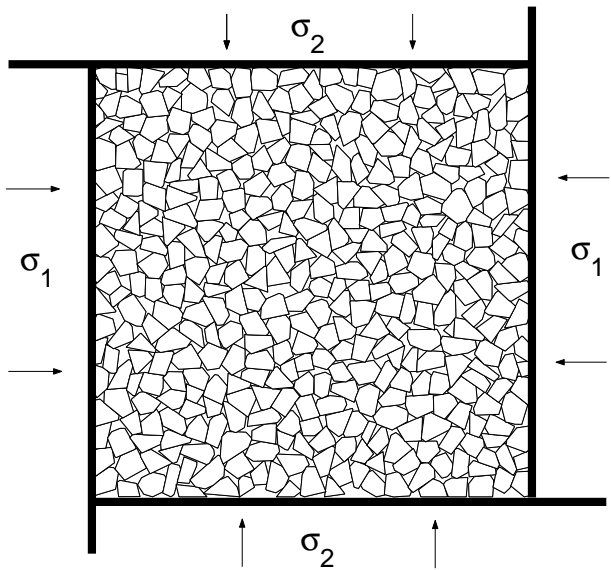,width=\linewidth}
    \caption{Polygonal assembly confined by a rectangular frame of walls.}
    \label{fig:sample}
  \end{center}
\end{figure}

The void ratio can be decreased by reducing the volume between 
the walls. The drawback of this method is that it leads to 
significant differences between the inner and outer parts 
of the boundary assembly, and it yields unrealistic overlaps 
between the particles, giving rise to enormous pressures. 
Alternatively, one can confine the polygons by applying gravity 
to the particles and on the walls. Particularly, homogeneous, 
isotropic assemblies, as shown in Fig. \ref{fig:sample} can be 
generated by a gravitational  field  $\vec{g}=-g\vec{r}$ 
where $\vec{r}$ is  the vector connecting the center of mass of  
the assembly to the center of the polygon. 

When the sample is consolidated, repeated shear stress cycles 
are applied from the walls, superimposed to a confining pressure. 
The external load is imposed by applying a force 
$[p_c+q_c\sin(2\pi t/t_0)]W$ and  $[p_c+q_c\cos(2\pi t/t_0)]H$ 
on the  horizontal and  vertical walls, respectively.  $W$ and $H$ 
are the width and the height of the sample. If we take $p_c=16kPa$ 
and $q_c < 0.4 p_c$ ,  we observe that the void ratio decreases as 
the number of cycles increases. Void ratios around $0.15$ can be 
obtained.  It is worth mentioning that after some thousands of cycles 
the void ratio is still slowly decreasing, making it difficult 
to identify this minimal value.

\section{Simulation results}
\label{simulation}

In order to investigate different aspects of the incremental response some 
numerical simulations were performed. Different polygonal assemblies 
of $400$ particles were used in the calculations. The loading between 
two stress states was controlled  by applying  forces 
$[\sigma^i_1+(\sigma^f_1-\sigma^i_1)r(t)]W$ and 
$[\sigma^i_2+(\sigma^f_2-\sigma^i_2)r(t)]H$.
A smooth modulation $r(t)=(1-\cos(2\pi t/t_0))/2$ is chosen in order to
minimize the acoustic waves produced  during loading. 
The initial void ratio is around $\nu=0.22$.

The components of the stress are represented by  $p=(\sigma_1+\sigma_2)/2$ 
and  $q=(\sigma_1-\sigma_2)/2$, where $\sigma_1$ and $\sigma_2$ are the 
eigenvalues of the averaged stress tensor on the RVE. Equivalently, the 
coordinates of the  strain are given by the  sum 
$\gamma=\epsilon_2-\epsilon_1$ and the  difference 
$e=-\epsilon_1-\epsilon_2$  of the eigenvalues of the strain tensor. 
We use the convention that $e>0$ means compression of the sample. 
The diameter of the RVE is taken $d=16\ell$ .
All the calculations were taken in the quasistatic regime.

\subsection{Superposition principle}
\label{superposition}

\begin{figure}[t]
\begin{center}
\epsfig{file=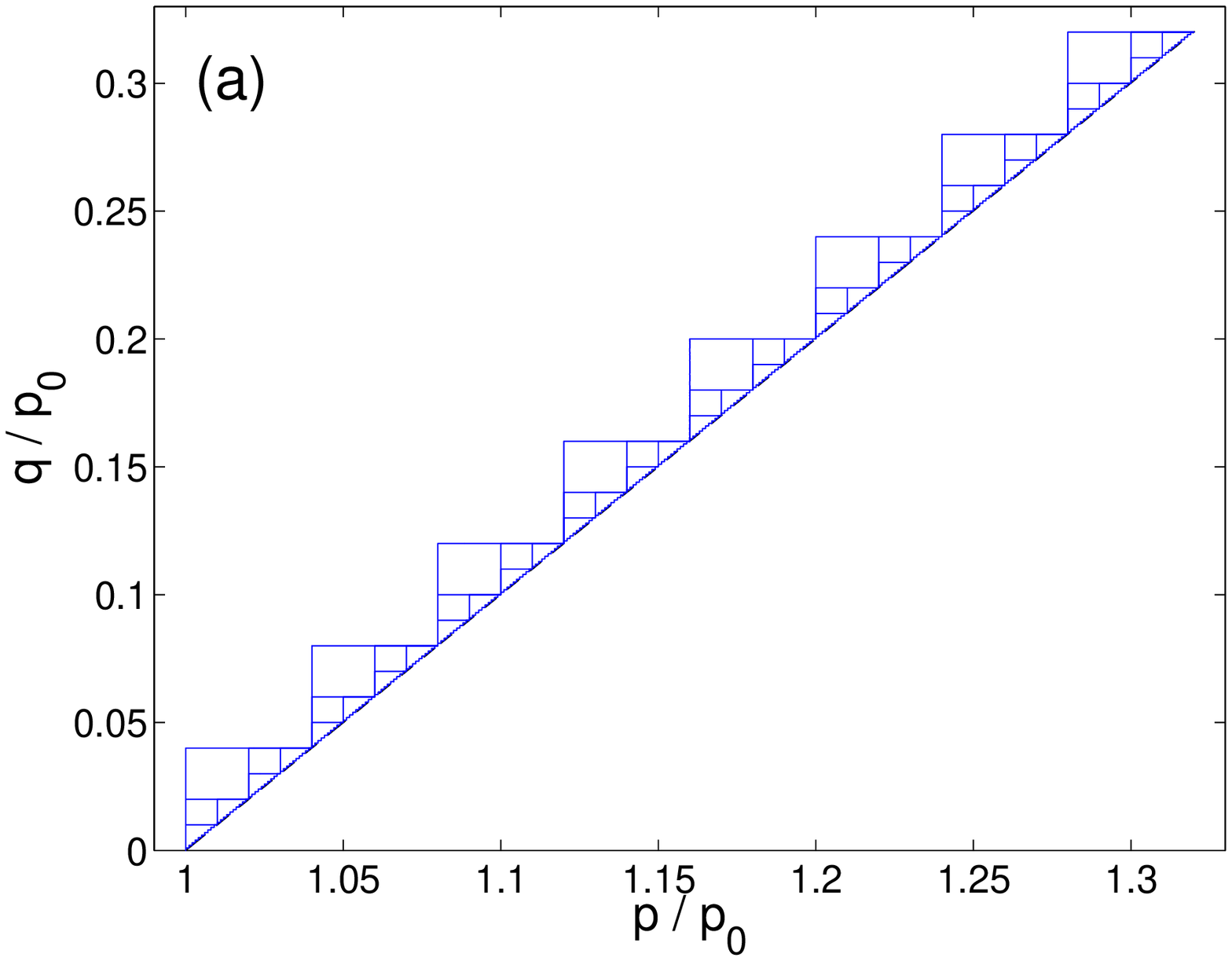,width=\linewidth,clip=1}
\epsfig{file=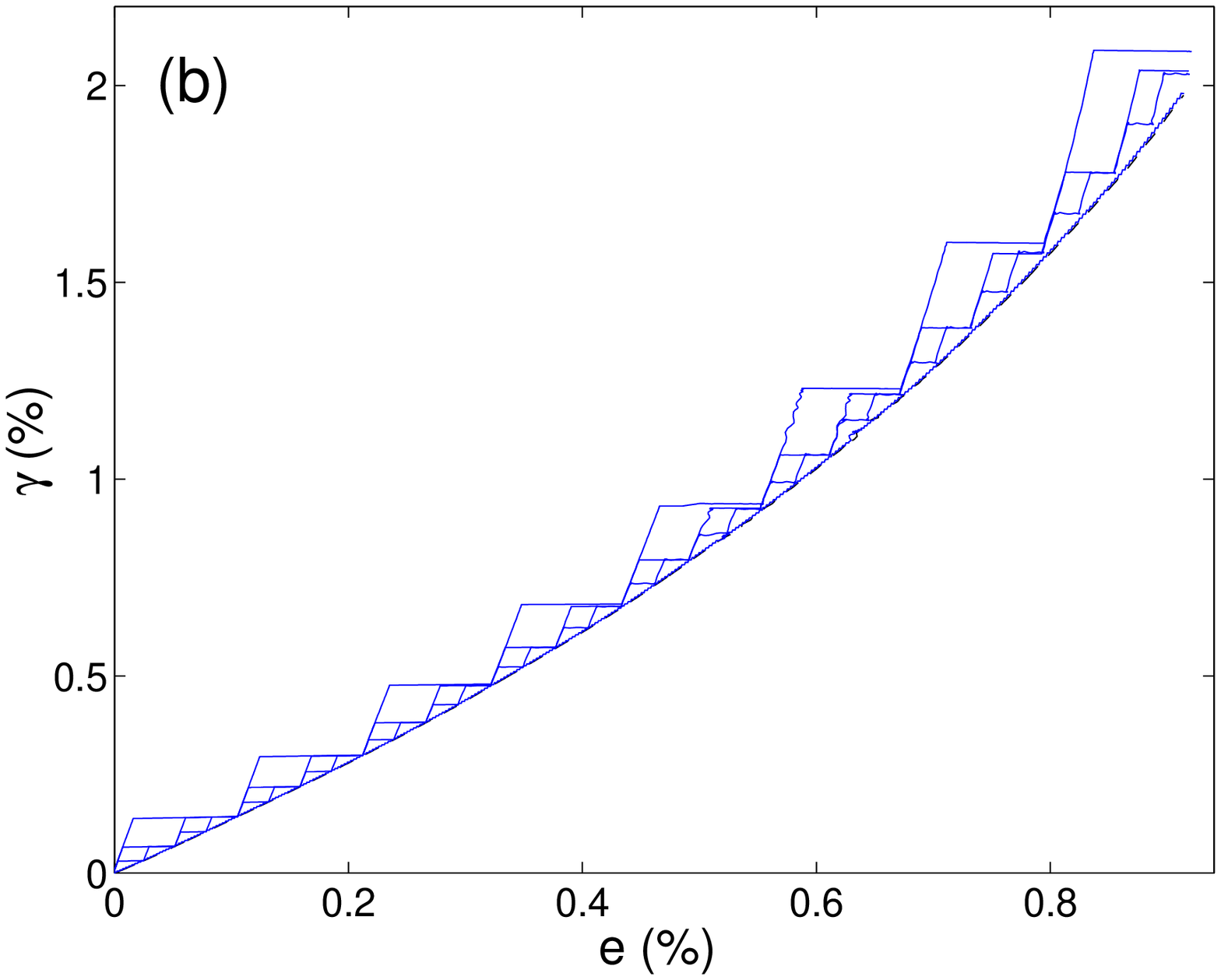,width=\linewidth,clip=1}
\caption{ Comparison between numerical responses obtained from MD 
simulations of a rectilinear proportional loading 
(with constant lateral pressure) and stair-like paths.}
\label{fig:superposition2}
\end{center}
\end{figure}

We start exploring the validity of the superposition principle 
presented in Subsec. \ref{hypoplastic}. The part (a) of Fig. 
\ref{fig:superposition2} shows the loading path during the 
proportional loading under constant lateral pressure. This path 
is also decomposed into pieces divided into two parts: one is an 
incremental isotropic loading  ($\Delta p = \Delta\sigma$ and 
$\Delta q = 0$), the other is an incremental pure shear loading  
($\Delta q = \Delta\sigma$ and $\Delta p = 0$). The length of the 
steps $\Delta\sigma$ varies from to $0.4p_0$ to $0.001p_0$, where 
$p_0=640kPa$.  The part (b) of Fig. \ref{fig:superposition2} shows 
that as the steps decrease, the strain  response converges to the 
response of the proportional loading. In order to verify this 
convergence, we calculate the difference between the strain response 
of the stair-like path $\gamma(e)$ and the proportional loading path 
$\gamma_0(e)$ as:

\begin{equation}
  \label{eq:difference}
  \Delta\epsilon \equiv \max_e{|\gamma(e)-\gamma_0(e)|},
\end{equation}

\begin{figure}[t]
  \begin{center}
    \epsfig{file=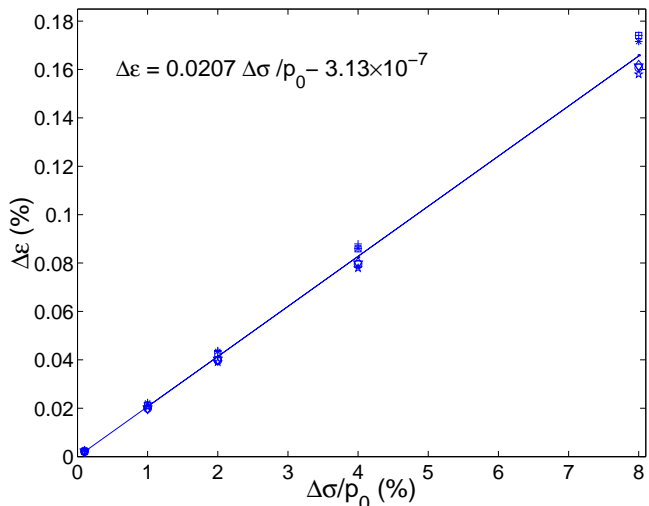,width=\linewidth,clip=1}
    \caption{Distance between the response of the stair-like path 
             and the proportional path.}
    \label{fig:superposition3}
  \end{center}
\end{figure}

\noindent
for different steps sizes. This is shown in Fig. \ref{fig:superposition3} 
for seven different polygonal assemblies.  The linear interpolation of 
this data intersects the vertical axis at $3\times 10^{-7}$. Since this 
value is below the error given by the quasi-static approximation, 
which is $3\times 10^{-4}$, the results suggest that the responses 
converge to that one of the proportional load. Therefore we find that
within our error bars the superposition principle is valid.

A close inspection of the incremental response will show that the validity 
of the superposition principle is linked to the existence of tensorial
zones in the incremental stress space. Before this, a short introduction 
to the strain envelope responses follows.

\subsection{Incremental response}
\label{incremental response}

A graphical illustration of the particular features of the constitutive 
models can be given by employing the so-called {\it response envelopes}. 
They were introduced by Gudehus \cite{gudehus79} as a useful tool to 
visualize the properties of a given incremental constitutive equation. 
The strain envelope response is defined as the image 
$ \{ d\tilde\epsilon={\cal G} (d\tilde{\sigma},\tilde{\sigma}) \} $ in 
the strain space of the unit sphere in the stress space, which becomes 
a potato-like surface in the strain space.

In practice, the determination of the stress envelope responses is 
difficult because it requires one to prepare many samples with 
identical material properties. Numerical simulations result 
as an alternative to the solution of this problem. They allow 
one to create clones of the same sample, and to perform different 
loading histories in each one of them.

In the case of a plane strain tests, where there is no deformation in 
one of the spatial directions, the strain envelope 
response can be represented in a plane.  According to Eq. 
(\ref{eq:hypoplastic1}), this response results in 
a rotated, translated ellipse in the hypoplastic theory, whereas 
it is given by a continuous curve consisting of two pieces of 
ellipses in the elasto-plastic theory, as result from Eq. 
(\ref{eq:elasto-plastic}). It is then of obvious 
interest to compare these predictions with the stress envelope 
response of the experimental tests.

\begin{figure}[t]
  \begin{center}
    \epsfig{file=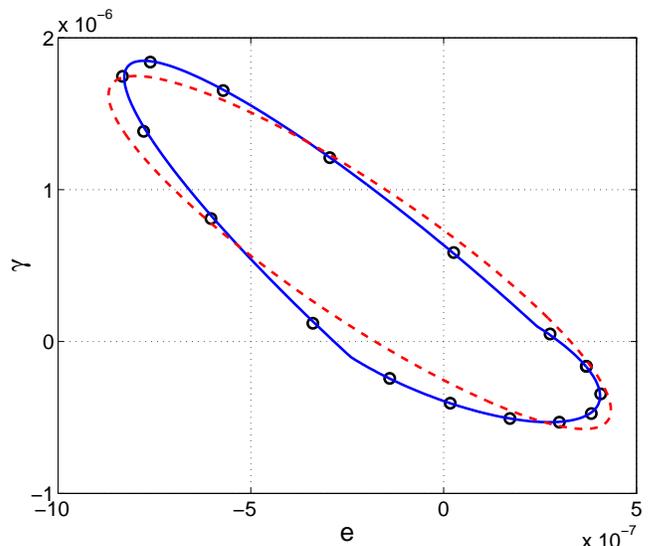,width=\linewidth,clip=1}
    \caption{Numerical calculation of the incremental strain response. 
The dots are the numerical results. The solid curve represents the 
fit to the elasto-plastic theory. The dashed curve is the hypoplastic 
fit.}
    \label{fig:hypo-elastic}
  \end{center}
\end{figure}

Fig. \ref{fig:hypo-elastic} shows the typical strain response 
resulting from the different stress controlled loading in a dense 
packing of polygons. Each point is obtained from the strain response 
in a specific direction of the stress space, with the same stress 
amplitude of $10^{-4} p_0$. We take $q_0=0.45p_0$ and $p_0=160kPa$
In this calculation. The best fit of these results 
with the envelopes response of the elasto-plasticity 
(two pieces of ellipses). For comparison the hypoplasticity (one ellipse) 
is also shown in Fig. \ref{fig:hypo-elastic}. 

From these results we conclude that the elasto-plastic theory 
is more accurate in describing the incremental response of our 
model. One can draw to the same conclusion taking different 
strain values with different initial stress values \cite{alonso02a}.
These results have shown that the incremental response can
be accurately described using the elasto-plastic relation of 
Eq. (\ref{eq:elasto-plastic}). The validity of this equation is 
supported by the existence of a well defined flow rule for each 
stress state.

\subsection{Yield function}
\label{yield funcion}

In Subsec. \ref{elasto-plastic}, we showed that the yield surface is
an essential element in the formulation of the Drucker-Prager
theory.  This surface encloses a hypothetical region in the 
stress space where only elastic deformations are possible 
\cite{drucker52}. The determination of such a yield surface is 
essential to determine the dependence of the strain response on 
the history of the deformation.

\begin{figure}[t]
 \begin{center}
 \epsfig{file=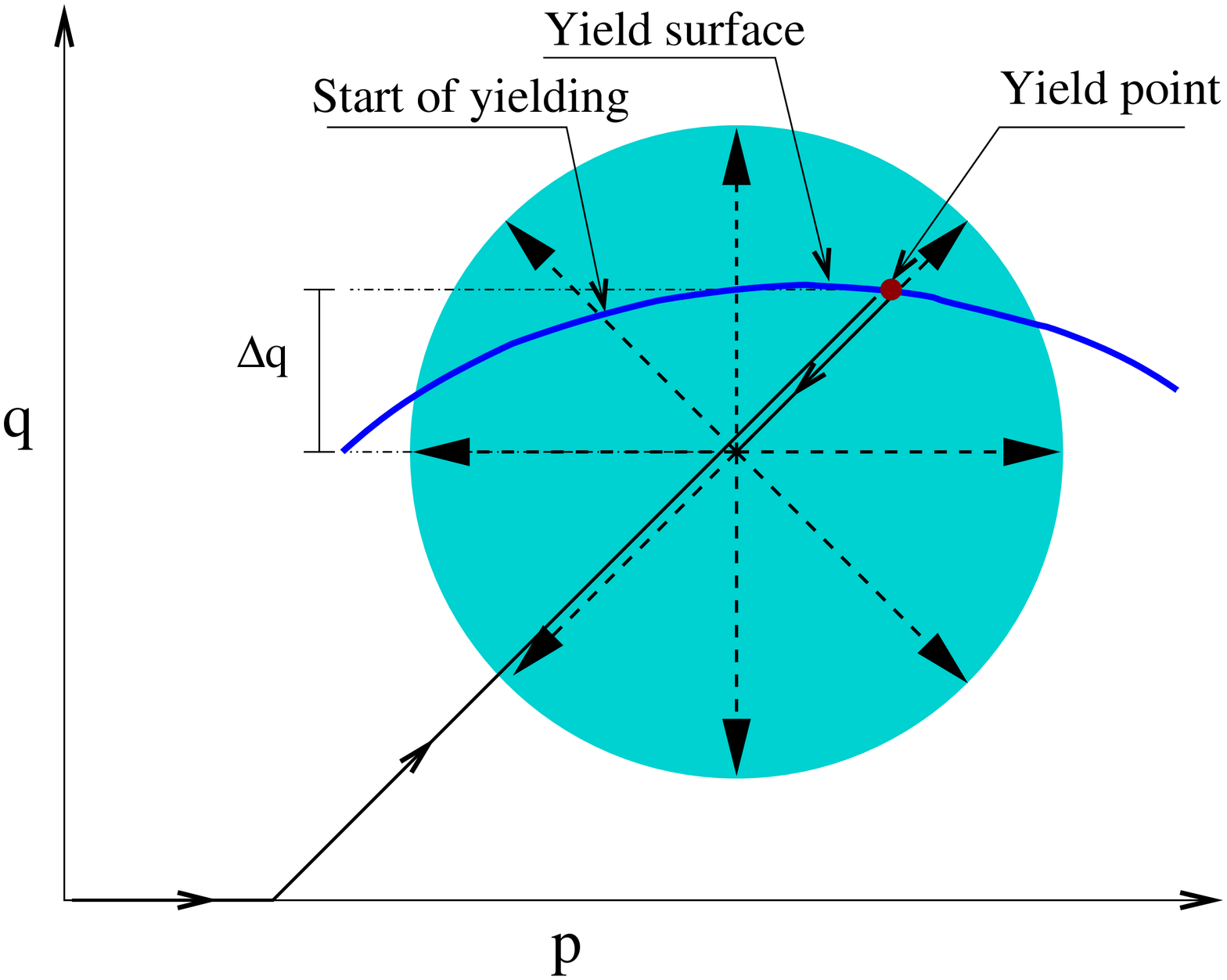,width=0.85
\linewidth,angle=0,clip=1}
 \end{center}
 \caption{Method to obtain the yield surface. Load-unload-reload tests are 
performed taking different directions in the reload path. In each direction, 
the point of the reload path where the yielding begins is marked. The yield 
function is constructed by connecting these points. }
 \label{yield}
\end{figure}

We attempt to detect the yield surface by using a standard procedure 
proposed in experiments with sand \cite{tatsouka74}. Fig.\ \ref{yield} 
shows this procedure. Initially the sample is subjected to an isotropic 
pressure. Then the sample is loaded in the axial direction until it 
reaches the yield-stress state with pressure $p$ and deviatoric stress 
$q$. Since plastic deformation is found at this stress value, the point 
$(p,q)$ can be considered as a classical yield point. Then, the 
Drucker-Prager theory assumes the existence of a yield surface 
containing this point. In order to explore the yield surface, 
the sample is unloaded in the axial direction until it reaches 
the stress point with pressure $p-\Delta p $ and deviatoric stress 
$q-\Delta p$ inside the elastic regime. Then the yield surface is 
constructed by re-loading in different directions in the stress space. 
In each direction, the new yield point must be detected 
by a sharp change of the slope in the stress-strain curve, indicating 
plastic deformations.

\begin{figure}[t]
 \begin{center}
 \epsfig{file=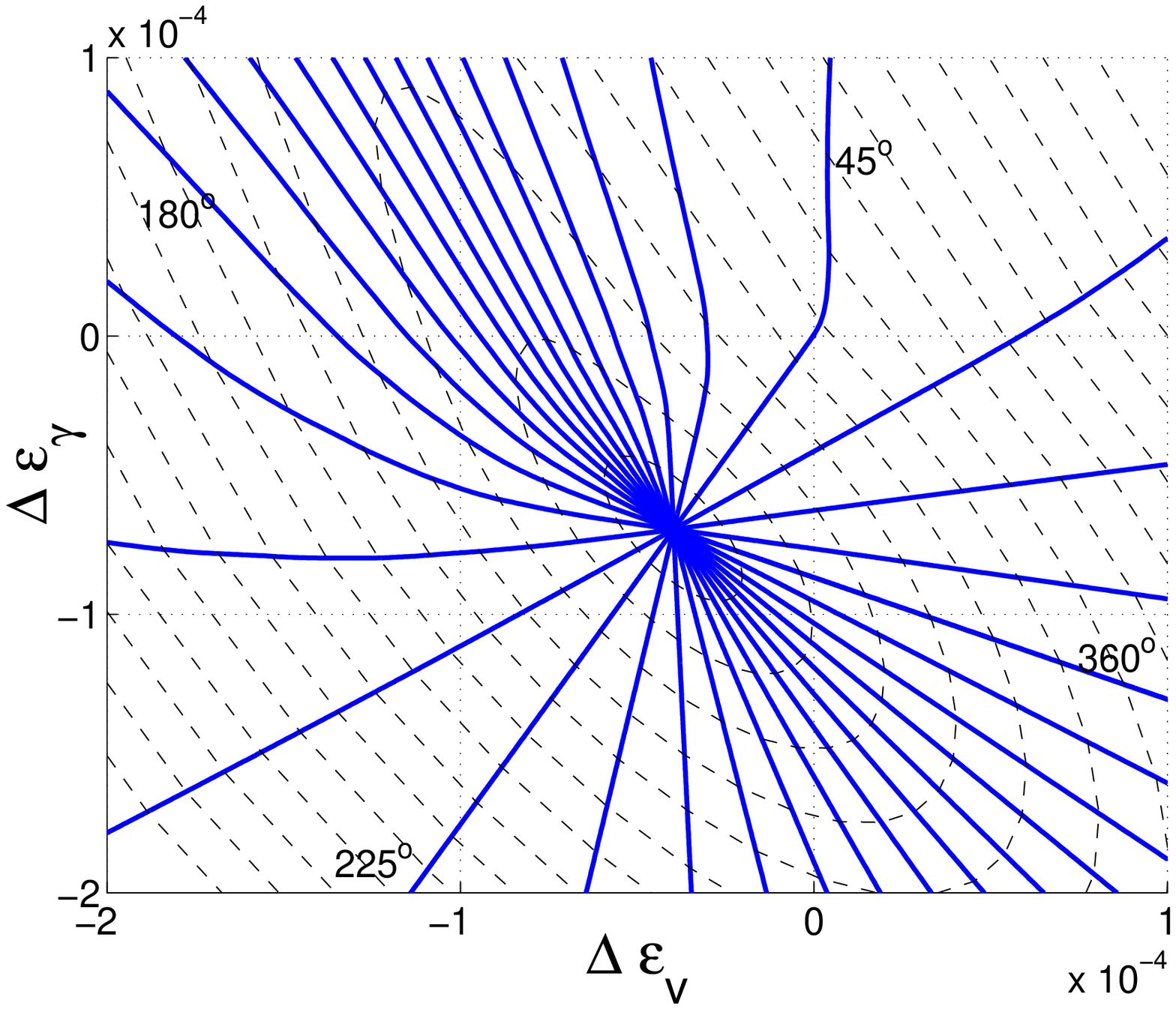,width=\linewidth,angle=0,clip=1}
 \end{center}
\caption{Strain responses according to Fig. \ref{yield}. The solid lines 
show the strain response from different reload directions. The dashed 
contours represent the strain envelope responses
for different stress increments $|\Delta\vec{\sigma}|$.}
\label{De}
\end{figure}

Fig.\ \ref{De} shows the strain response taking different load 
directions in the same sample. The initial stress of the sample
is given by $q_0=0.5p_0$ and $p_0=160kPa$. If the direction of the reload 
path is the same as that of the original load ($45^o$), we 
observe a sharp decrease of stiffness when the load point 
reaches the initial yield point, which corresponds to the 
origin in Fig.\ \ref{De}. However, if one takes a direction 
of re-loading different from $45^o$, the decrease of the 
stiffness with the loading becomes smooth. Since there 
is no straightforward way to identify those points where 
the yielding begins, the yield function, as it was 
introduced by Drucker and Prager \cite{drucker52} in 
order to describe a  sharp transition between the elastic 
and plastic regions, is not consistent with our results.

Experimental studies on dry sand seem to show that the truly 
elastic region is probably extremely small \cite{gudehus84}.
Moreover, a micro-mechanic investigation of the mechanical 
response of granular ratcheting under cyclic loading  has
shown that any load involves sliding contacts, and hence, 
plastic deformation \cite{alonso04}. These studies draw to 
the conclusion that the elastic region, used in the Drucker-Prager 
theory to give a dependence of the response on recent history, 
is not a necessary feature of granular materials. 

A question that naturally rises is if the hypoplastic theory is 
more appropriate than the elasto-plastic models to describe soil
plasticity. Since these models do not introduce any elastic
regime, they seem to provide a good alternative. However, the
modern versions of hypoplasticity depart from the superposition 
principle, which is not consistent with our results.
An alternative approach to hypoplasticity can be reached from 
the bounding surface theory,  by shrinking the  elastic regime 
to the current stress point \cite{dafalias86a}. With this limit 
one can reproduce  the  observed continuous  transition from the 
elastic to the  elasto-plastic behavior and in the same time keep 
the tensorial zones. However, it has been shown that this limit 
leads to a constitutive relation in terms of some internal 
variables, which lack of physical meaning in this theory. In the 
author's opinion, the necessity to provide a micro-mechanical 
interpretation of these  internal variables will be important to 
capture this essential feature of  mechanics of granular materials, 
that any loading involves plastic deformation.

\section{Instabilities}
\label{instabilities}

Instability has been one of the classical topics of soil 
mechanics. Localization from a previously homogeneous 
deformation to a narrow zone of intense shear is a common 
mode of failure of soils \cite{roscoe70,vermeer84,desrues84}. 
The Mohr-Coulomb criterion is typically used to 
understand the principal features of the localization. 
This criterion was improved by the Drucker condition, based 
on the hypothesis of the normality, which results in a plastic 
flow perpendicular to the yield surface \cite{drucker52}. This 
theory predicts that the instability appears when the stress of 
the sample reaches the plastic limit surface. This surface is 
given by the stress states where the plastic deformation becomes 
infinite. In previous work, it is shown that the normality postulate 
is not fulfilled in the incremental response of this model, because
the flow and yield directions of Eq. (\ref{eq:plastic-flow2})
are different \cite{alonso02a}.  Thus, it is interesting to see if the 
Drucker stability  criterion is still valid.

According to the flow rule of Eq. (\ref{eq:plastic-flow2}), the plastic 
limit surface can be found by looking for the stress values 
where the plastic modulus vanishes. The dependence of this 
modulus on the stress fits to the following power law relation 
\cite{alonso02a}: 

\begin{equation}
h= h_0 \left[ 1 - \frac{q}{q_0}(\frac{p_0}{p})^{\vartheta}\right]^{\eta}.
\label{hardening2} 
\end{equation}

\noindent
This is given in terms of the four parameters: The plastic modulus 
$h_0 = 14.5 \pm 0.05$ at $q=0$, the constant $q_0=0.85 \pm 0.05$, 
and the exponents $\eta = 2.7 \pm 0.04 $ and  $\vartheta =  0.99 \pm 0.02$.  
Then, the plastic limit surface is given by the stress states with zero 
plastic modulus: 

\begin{equation}
\frac{q_p}{q_0}= \left(\frac{p}{p_0}\right)^{\vartheta}. 
\label{pline}
\end{equation}

On the other hand,  the failure surface can be defined by the limit
of the stress values where the material becomes unstable. It has been 
shown that this is given by the following relation \cite{alonso02a}

\begin{equation}
\frac{q}{q_c}=\left(\frac{p}{p_0}\right)^{\beta}. 
\label{fline}
\end{equation}

\noindent
Here $p_0=1.0MPa$ is the reference pressure, and $q_c=0.78\pm0.03 MPa$. 
The power law dependence on the pressure, with exponent 
$\beta = 0.92 \pm 0.02 $  implies a small deviation from the 
Mohr-Coulomb theory where the relation is strictly linear.

By comparing Eq. (\ref{fline}) to Eq. (\ref{pline}) one observes that 
during loading the instabilities appear before reaching the plastic 
limit surface. Theoretical studies have also shown that in the case of 
non-associated materials, (i.e. where flow direction does not agree with 
the yield direction) the instabilities can appear strictly inside of the 
plastic limit surface \cite{darve00}. In this context, the question of 
instability must be reconsidered beyond the Drucker condition.  

The stability for non-associated elasto-plastic materials has been 
investigated by Hill, who established the following sufficient 
stability criterion  \cite{hill58}.

\begin{equation}
  \label{eq:hills}
  \forall d\tilde\epsilon,~~~ d\tilde\sigma \cdot d\tilde\epsilon > 0.
\end{equation}
  
The analysis of this criterion of stability will be presented here based 
on the constitutive relation given by Eq. (\ref{eq:elasto-plastic}). 
The stability condition of this constitutive relation can be evaluated 
by introducing the normalized second order work \cite{darve00}:

\begin{equation}
  d^2W \equiv  \frac{d\tilde\sigma \cdot d\tilde\epsilon}{|d\tilde\sigma|^2}
\label{eq:d2W}
\end{equation}

Then, the Hill condition of stability can be obtained by replacing 
Eq. (\ref{eq:elasto-plastic}) in this expression. This results in

\begin{equation}
  d^2W = \hat\sigma D^{-1} \hat\sigma + \frac{\langle\cos(\theta+\phi)\rangle}{h}\cos(\theta+\psi)>0,
\label{eq:d2W2}
\end{equation}

\noindent
where $\hat\sigma$ was defined in Eq. (\ref{dir}).
In the case where the  Drucker normality postulate $\psi=\phi$ is valid, 
Eq. (\ref{eq:d2W2}) is strictly positive and, therefore,
this stability  condition would be valid for all the stress states inside 
the plastic limit surface . On the contrary, for a non-associated flow rule as 
in our model, the second order work is not strictly positive for all 
the load directions, and it can take zero values inside the plastic 
limit surface (i.e. during the hardening regime where $h>0$).

\begin{figure}[t]
  \begin{center}
     \epsfig{file=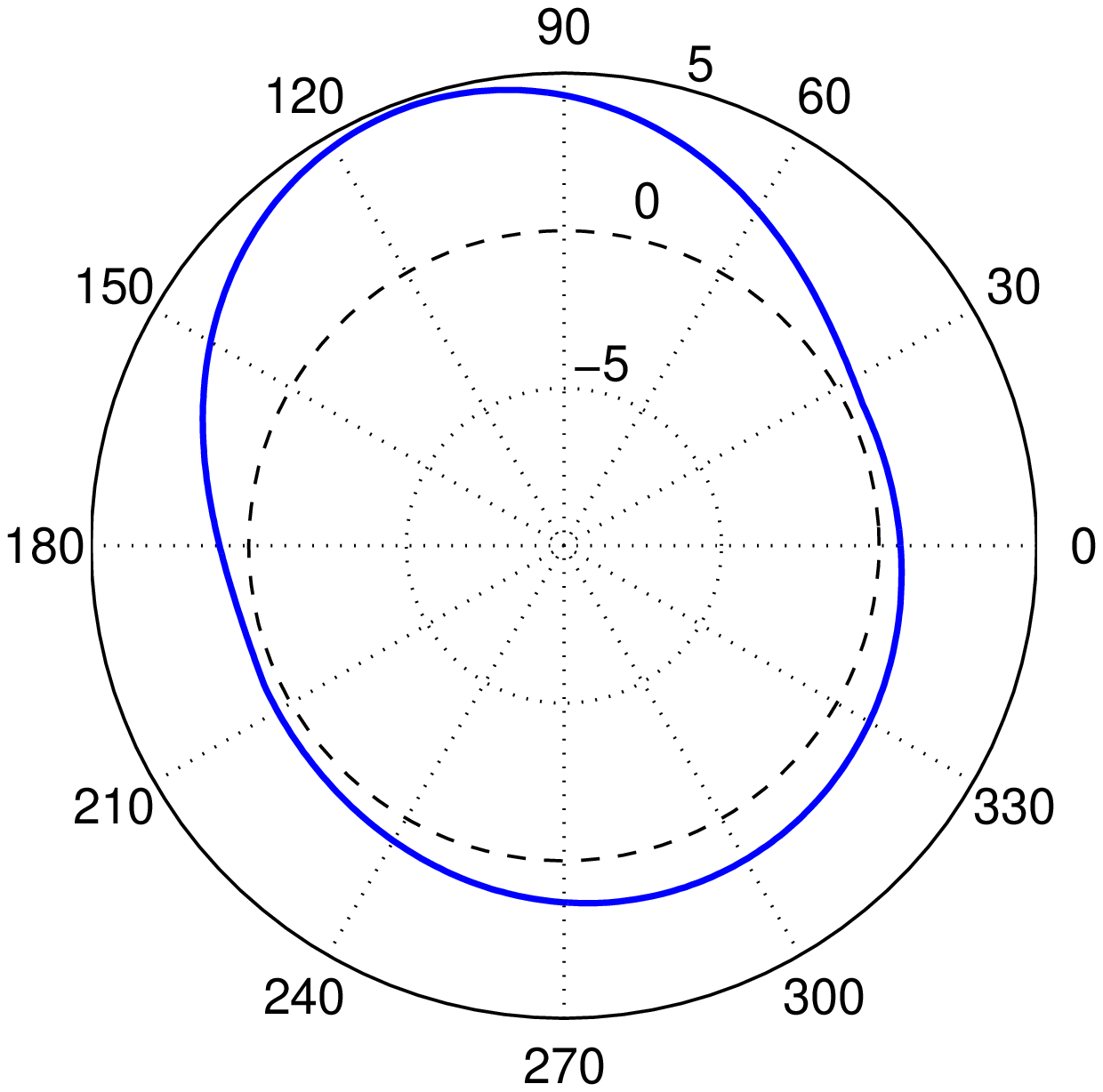,width=0.3\linewidth}~~~
     \epsfig{file=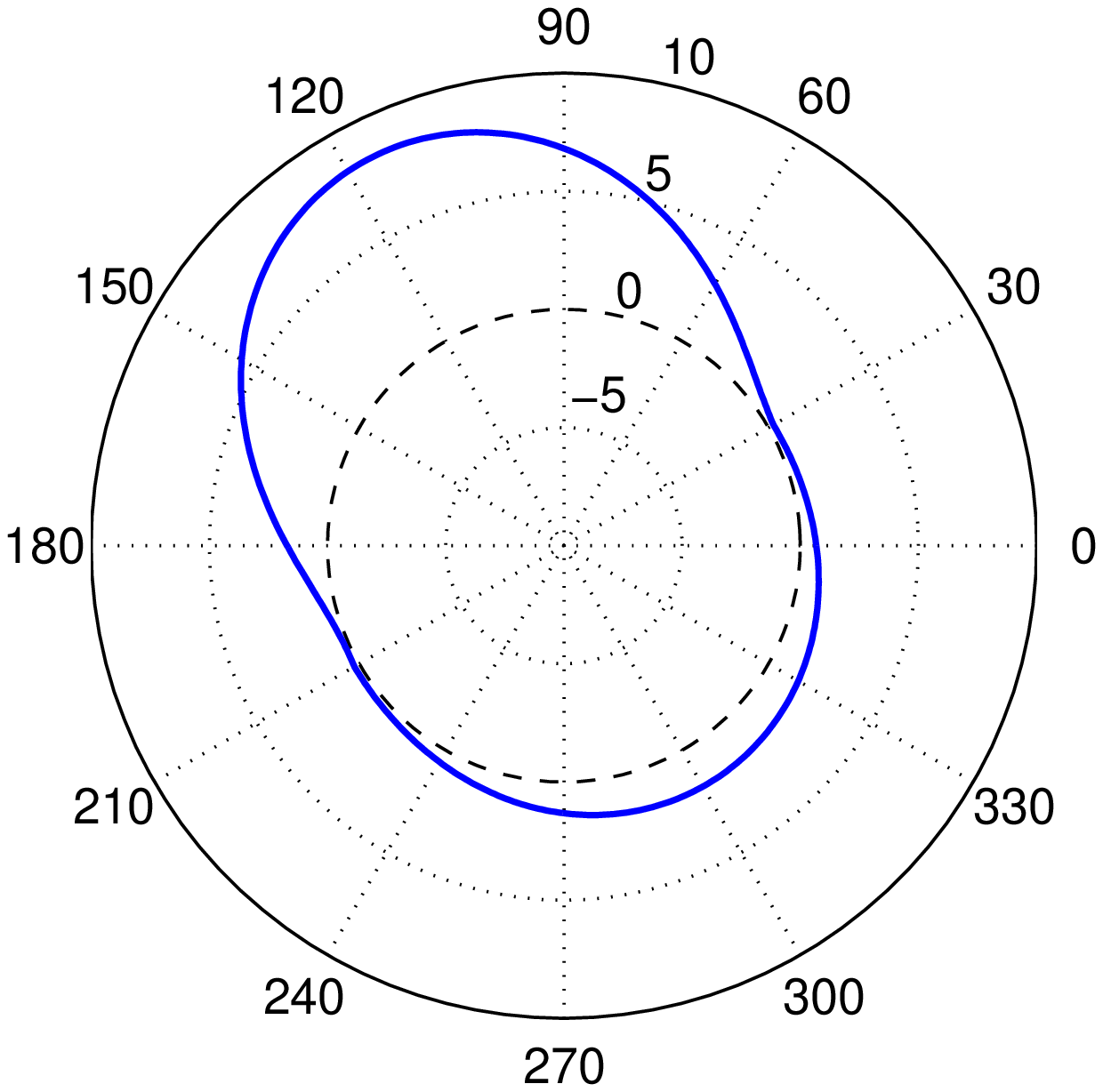,width=0.3\linewidth}~~~
     \epsfig{file=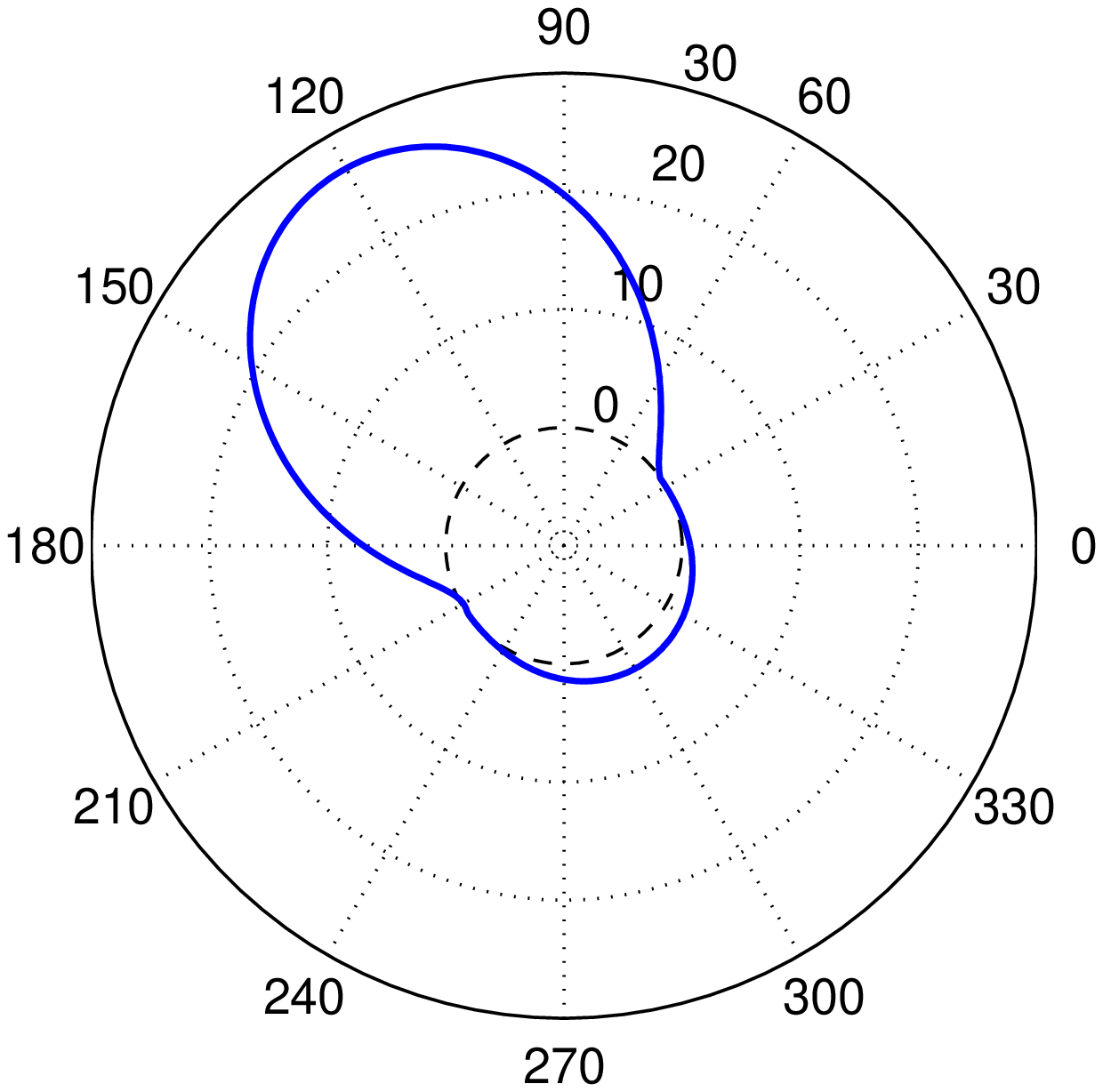,width=0.3\linewidth}~~~
    \caption{The solid lines show the second order work as a 
function of the direction of load for three different stress ratios  
$q/p=0.6$ (left), $0.7$ (center), and $0.65$ (right) with pressure 
$p=160KPa$. The dashed circles enclose the region where $d^2W<0$.  }
    \label{fig:d2W}           
  \end{center}
\end{figure}

To analyze this instability, we adopt a circular representation 
of $d^2W$ shown in Fig. \ref{fig:d2W}. The dashed circles in these figures 
represent those regions where $d^2W<0$. For stress ratios below $q/p=0.7$ 
we found that the second order work is strictly positive. Thus, 
according to the Hill stability condition, this region corresponds 
to stable states. For the stress ratio $q/p=0.8$, the second order 
work becomes negative between $27^o<\theta<36^o$ and $206^o<\theta<225^o$. 
This leads to a domain of the stress space strictly inside the plastic 
limit surface where the Hill condition of stability is not fulfilled,
and therefore the material is potentially unstable.

Numerical simulations of biaxial tests show that strain localization 
is the most typical mode of failure \cite{cundall82,astrom00}. The fact 
that it appears before the sample reaches the plastic limit surface suggests 
that the appearance of this instability is not completely determined by the 
current macroscopic stress of the material, as it is predicted by the 
Drucker-Prager theory.  More recent analytic \cite{muhlhaus87b} and 
experimental \cite{desrues84,marcher01} works have focused on the role of the 
micro-structure on the localized instabilities. Frictional slips at 
the particles have been used to define additional degrees of freedom 
\cite{muhlhaus87b}. The introduction of the particle diameter in the 
constitutive relations results in a correct prediction of the shear band 
thickness. The new degrees of freedom of these constitutive models are 
still not clearly connected to the micro-mechanical variables of 
the grains, but with the development of numerical simulations
this aspect can be better understood.

\end{multicols}

\begin{multicols}{2}

\section{Concluding remarks}

In this paper we have obtained explicit expressions for the 
averaged stress and strain tensors over a RVE, in terms of the 
micro-mechanical variables, contact forces and the individual 
displacements and rotations of the grains. 

A short review on the incremental non-linear models
has been presented. We emphasize  the existence of the
elastic regime, and the two tensorial zones as predicts the
theory of elasto-plasticity. We consider also the 
superposition principle of soil mechanics, which is not
satisfied in the incremental non-linear models. These assumptions
have been studied using molecular dynamics simulations on a 
polygonal packing. The results are summarized as follows:

\begin{itemize}

\item The elasto-plastic theory, with two  tensorial zones, provided a 
more accurate description of the incremental  response than the 
hypoplastic theory.

\item  In contradiction to the incremental non-linear models, the 
simulation results  show that the superposition principle is accurately 
satisfied. 

\item The experimental method proposed by Tatsouka has been implemented to 
identify the yield surface. The resulting strain response shows that the 
transition from elasticity to elasto-plasticity is not as sharp as the 
Drucker-Prager theory predicts, but a smooth transition occurs. The fact 
that there is no purely elastic regime leads to the open question of how 
to determine the dependence of the response of soils on the history of 
the deformation.

\item The calculation of the plastic limit condition and the failure surface 
shows that the failure appears during the hardening regime $h>0$. This result 
is analyzed using the Hill condition of stability, which states that for 
non-associated materials the instabilities can appear before the plastic 
limit surface.

\end{itemize}

These conclusions appear to contradict both the Drucker-Prager theory 
and the hypoplastic models. In future work, it would be important to revisit 
the question of the incremental non-linearity of soils from micro-mechanical 
considerations.

\section*{Acknowledgments}

We thank F. Darve, Y. Kishino, D. Kolymbas, F. Calvetti, 
Y.F. Dafalias S. McNamara and R. Chambon for helpful discussions 
and acknowledge 
the support of the  {\it Deutsche Forschungsgemeinschaft\/} within the 
research group   {\it Modellierung koh\"asiver Reibungsmaterialen\/} 
and the European DIGA project HPRN-CT-2002-00220.


\bibliographystyle{unsrt}


\end{multicols}

\end{document}